\documentclass[pra,twocolumn,superscriptaddress,showpacs,amsmath,amstex,amssymb,citeautoscript]{revtex4-1}

\usepackage[T1]{fontenc}
\usepackage[utf8]{inputenc}
\usepackage{lipsum}
\usepackage{amsmath}
\usepackage{amssymb}
\usepackage{bm}
\usepackage{bbm}
\usepackage{braket}
\usepackage{xcolor}
\usepackage{pifont}
\usepackage[mathscr]{euscript}
\usepackage[shortlabels]{enumitem}
\usepackage[justification=justified]{subcaption}
\usepackage{graphicx}
\usepackage{lipsum}

\usepackage{graphicx}
\usepackage{subcaption}
\allowdisplaybreaks
\usepackage{float}
\usepackage{graphicx}
\usepackage{dsfont}
\usepackage{comment}
\usepackage[colorlinks=true]{hyperref}  
\hypersetup{
    bookmarks=true,         
    unicode=false,          
    pdftoolbar=true,        
    pdfmenubar=true,        
    pdffitwindow=false,     
    pdfstartview={FitH},    
    pdftitle={Superconductivity in rhombohedral graphene},    
    pdfauthor={Christos, Bonetti, Scheurer},     
    pdfsubject={},   
    pdfcreator={},   
    pdfproducer={}, 
    pdfkeywords={} {} {}, 
    pdfnewwindow=true,      
    colorlinks=true,       
    linkcolor=blue, 
    citecolor=blue,        
    filecolor=magenta,      
    urlcolor=blue           
}

\newcommand{\equref}[1]{Eq.~(\ref{#1})}

\newcommand{\figref}[1]{Fig.~\ref{#1}}
\newcommand{\refcite}[1]{Ref.~\onlinecite{#1}}

\newcommand{\diff}{\mathrm{d}}
\newcommand{\sign}{\,\text{sign}}

\renewcommand{\Re}{\text{Re}}
\renewcommand{\Im}{\text{Im}}

\renewcommand{\vec}[1]{\boldsymbol{#1}}
\newcommand{\eo}{\xi_{\vec{k}}}
\newcommand{\etw}{\xi_{\vec{q}_1-\vec{k}}}
\renewcommand{\eth}{\xi_{\vec{q}_2-\vec{k}}}

\definecolor{wrongultramarine}{rgb}{1,0.5,0}

\begin{document}

\title{Finite-momentum pairing and superlattice superconductivity \\ in valley-imbalanced rhombohedral graphene}

\author{Maine Christos}
\affiliation{Department of Physics, Harvard University, Cambridge MA 02138, USA}

\author{Pietro M.~Bonetti}
\affiliation{Department of Physics, Harvard University, Cambridge MA 02138, USA}

\author{Mathias S.~Scheurer}
\affiliation{Institute for Theoretical Physics III, University of Stuttgart, 70550 Stuttgart, Germany}

\begin{abstract}
Inspired by the recent experimental discovery of superconductivity emerging from a time-reversal symmetry-breaking normal state in tetralayer rhombohedral graphene, we here  investigate superconducting instabilities in this system. We classify the possible pairing instabilities, including states with commensurate and incommensurate center of mass momenta. As rotational symmetry is broken in the latter type of pairing states, their momentum-space structure is most naturally characterized by a ``valley-independent Chern number’’, measuring the relative chirality between the normal and superconducting state. We further demonstrate that superconductivity can condense at multiple incommensurate momenta simultaneously, leading to the spontaneous formation of a translational-symmetry-breaking superlattice superconductor. Studying multiple different pairing mechanisms and varying the degree of spin and valley polarization in the normal state, we compare the energetics of these superconductors. Our results demonstrate that valley-imbalanced rhombohedral tetralayer graphene can give rise to rich superconducting phenomenologies.
\end{abstract}

\maketitle

Not long after the discovery of superconductivity, correlated insulating, and symmetry-broken metallic phases in graphene-based moiré superlattices \cite{review1,review2,review3}, it was shown that also non-twisted rhombohedral stacks of graphene layers can give rise to similar correlated physics \cite{SCTrilayer,HalfQuarterMetals,TwistedNonTwisted,PhysRevLett.127.187001,PhysRevLett.127.247001,PhysRevB.105.134524,ShubhayuPairing,PhysRevB.107.104502,PhysRevB.107.L161106,PhysRevB.105.L081407,PhysRevLett.130.146001,PhysRevB.108.045404,PhysRevB.108.134503,2024arXiv240617036D}. Recently, the discovery of superconductivity emerging from a valley-imbalanced normal state in rhombohedral-stacked tetralayer graphene \cite{Han2024a} has sparked a surge of interest \cite{qin2024chiralfinitemomentumsuperconductivitytetralayer,sau2024theoryanomaloushalleffect,yang2024topologicalincommensuratefuldeferrelllarkinovchinnikovsuperconductor,geier2024chiraltopologicalsuperconductivityisospin,chou2024intravalleyspinpolarizedsuperconductivityrhombohedral,2025arXiv250219474P,2024arXiv241109664J,2025arXiv250217555Y,2025arXiv250219474P,Kim_2025,PhysRevResearch.6.043240,2025arXiv250309692B,TheSTMPaper}. 

Superconductivity in the presence of valley imbalance is remarkable since time-reversal symmetry (TRS) is broken---a key symmetry for superconductors, as it guarantees the degeneracy of Kramers partners forming Cooper pairs. As such, Cooper pairs are no longer expected to always have vanishing center of mass (COM) momentum~\cite{ZHan2022} and the superconductor is predicted to become non-reciprocal, i.e., display a superconducting diode effect \cite{scammell_theory_2022}. Furthermore, as the normal state is already chiral itself, the superconductor is by construction chiral too. Nonetheless, it is still a meaningful and important question to clarify how this chirality affects the orbital character of the superconducting state and how superconducting instabilities can be classified in light of the reduced symmetries. In analogy to topological superconductivity induced by coupling $s$-wave superconductors to chiral systems \cite{Alicea_2012}, one might wonder whether the normal state’s Berry curvature will always be inherited by the superconductor, making it topologically non-trivial, or whether it depends on microscopic energetic details.

In this work, we address these questions by a combination of symmetry-based classification of pairing and energetics. Using the continuum model \cite{Zhang_2010,PhysRevB.80.165409} of rhombohedral tetralayer graphene and multiple different interactions mediated by fluctuations of bosonic modes, including phonons and spin fluctuations, we find that sufficiently large valley polarization not only leads to intravalley pairing (commensurate center-of-mass momentum $2\vec{K}$), as expected, but also to pairing at finite momentum $\vec{q}$ within the active valley (incommensurate net Cooper-pair momenta $2\vec{K} + \vec{q}$). While, at finite $\vec{q}$, one cannot use rotational symmetry to classify the orbital structure of the order parameter, we introduce a ``valley-independent Chern number” $\bar{C}$, where $|\bar{C}| \neq 0$ corresponds to a topologically non-trivial superconducting state and the sign of $\bar{C}$ encodes the relative chirality between the normal and superconducting phase. We find that, depending on the pairing mechanism and the normal state’s spin splitting, both topologically trivial and non-trivial phases are possible; for the latter, the chirality of the superconductor is found to be of the same orientation as that of the normal state ($\bar{C} > 0$) for all interactions we consider. 
Another important aspect is whether the order parameter condenses at a single wave vector $\vec{q}$ (which we call a 1-$\vec{q}$ state) or three $C_{3z}$-related wave vectors (3-$\vec{q}$ state). We find that microscopic parameters in the underlying continuum model of the theory determine which scenario is energetically favored. While the 1-$\vec{q}$ state preserves translational symmetry in observable quantities, the 3-$\vec{q}$ state breaks it, leading to a superconducting superlattice. This can be detected in STM experiments \cite{TheSTMPaper} and is mathematically described in analogy to the continuum model of moiré graphene \cite{dos2007graphene,bistritzer2011moire}, with interlayer terms corresponding to particle-hole couplings.


\begin{figure}
    \centering
    \includegraphics[width=\linewidth]{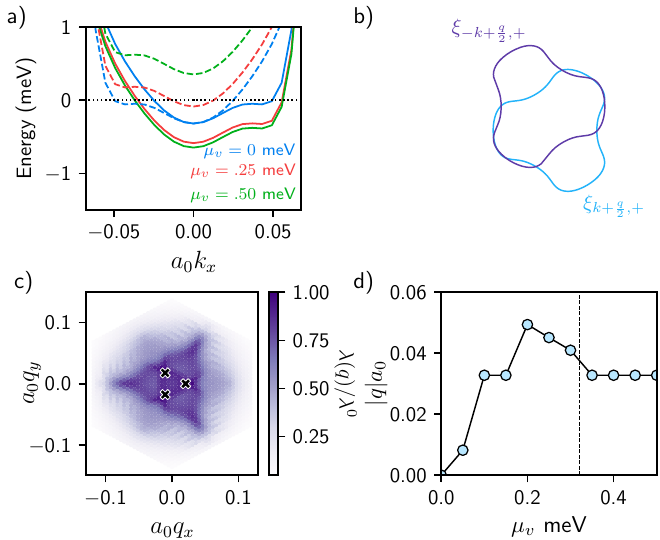}
    \caption{Band structure (a) for different $\mu_v$ in $+$ valley (solid lines) and $-$ valley (dashed lines). In b), we illustrate the energetically favorable value of $\vec{q}$ for a valley-polarized Fermi surface, leading to the maximum approximate overlap of the Fermi surfaces of $\xi_{\vec{k}+\frac{\vec{q}}{2},\eta,\sigma}$ and $\xi_{-\vec{k}+\frac{\vec{q}}{2},\eta,\sigma}$. (c) Eigenvalue of the leading instability $\lambda(\vec{q})$ normalized by its maximum $\lambda_0$ for ferromagnetic fluctuations as a function of center-of-mass momentum $\vec{q}$ for a fully valley-polarized and spin-degenerate normal state with global maxima marked by $\times$. In d), we plot the value of the Cooper pair center of mass momentum $\vec{q}$ as a function of the amount of valley polarization $\mu_v$ for $\mu_s=0$. The dashed black line denotes where the normal state becomes fully valley polarized. }
    \label{fig:fig1}
\end{figure}

\vspace{1em}
\textit{Model}---We model the normal state for superconductivity with the single-particle Hamiltonian:
\begin{equation}\label{Eq:kineticterm}
    \mathcal{H}_{c}=\sum_{\vec{k}}\sum_{\eta,\sigma}c^\dagger_{\vec{k},\eta,\sigma}\left(\epsilon_{\vec{k},\eta}-\mu+\eta\mu_{v}+\sigma\mu_{s}\right)c_{\vec{k},\eta,\sigma}
\end{equation}
In the above, $\epsilon_{\vec{k},\eta}$ is the dispersion of the active band of the continuum model of tetralayer rhombohedral graphene \cite{Zhang_2010,PhysRevB.80.165409,PhysRevB.107.104502} at small, positive electron density and $\eta$ and $\sigma$ are summed over valley and spin indices, respectively. In all our numerics, we compute $\epsilon_{\vec{k},\eta}$ with the same model parameters given in \cite{PhysRevB.107.104502}. Further discussion of this continuum model description and the model parameters used can be found in App.~\ref{App:LGE}. We model spin and valley polarization in the normal state with $\mu_s$ and $\mu_v$, which phenomenologically capture at a mean-field level the iso-spin symmetry breaking induced by electronic interactions at energy scales larger than those of the superconductor. In general, both $\mu_s$ and $\mu_v$ are allowed to have momentum dependence, though we choose them to be momentum indepent in our analysis. As these parameters are very difficult to predict theoretically or extract from experiment, we will study the superconducting phase diagram as a function of them. The chemical potential $\mu$ is always adjusted such that the electron density is fixed to $n_e=0.6 \times 10^{12}$ cm$^{-2}$, such that the density is near the value where valley-imbalanced superconductivity was observed in \cite{Han2024a}. The band structure of \equref{Eq:kineticterm} is plotted in Fig.~\ref{fig:fig1}a for different values of $\mu_v$.

The model we consider has an SU(2)$_s$ spin rotation symmetry when $\mu_s=0$, (spinless) TRS relating the two valleys when $\mu_v=0$, and U(1) valley symmetry for arbitrary $\mu_{s,v}$. Additionally, \equref{Eq:kineticterm} is invariant under $C_{3v}$ point group symmetries, generated by in-plane reflection $\sigma_v$ and threefold rotation $C_{3z}$ about the $z$ axis. As will become important for our discussion of intravalley pairing below, we will choose the $C_{3z}$ rotation axis to go through the $A_1$ sublattice and further take the phase of its representation in the microscopic 8-band model such that its $A_1$-$A_1$ matrix element is $1$ (see App.~\ref{App:LGE}).

We study interactions which originate from coupling the low-energy electrons to collective bosonic modes $\phi_j(\vec{q})=\phi^\dagger_j(-\vec{q})$ with coupling of the form:
\begin{equation}\label{Eq:fbcoupling}
    \mathcal{H}_{\text{fb}}=\sum_{\vec{k},\vec{p}}\sum_{\eta,\eta',\sigma,\sigma'}\phi_j(\vec{p})c^\dagger_{\vec{k},\eta,\sigma}F_{\vec{k},\vec{p},j}^{\eta\sigma,\eta'\sigma'}c_{\vec{k}+\vec{p},\eta',\sigma'}
\end{equation}
Such interactions can be thought of as originating from fluctuations of the various possible types of iso-spin polarization which may be proximate to superconductivity in the system, in analogy to the spin-fermion models initially studied in the context of high-temperature superconductivity \cite{abanov2001quantumcriticaltheoryspinfermionmodel}. We impose the aforementioned symmetries also on the entire interacting Hamiltonian such that the transformation properties of the bosons $\phi_j$ under these symmetries are fully determined by the form factors (no summation convention) $F_{\vec{k},\vec{q},j}^{\eta\sigma,\eta'\sigma'}=\rho^{\eta\sigma,\eta'\sigma'}_j\langle u_{\eta\sigma}(\vec{k})|u_{\eta'\sigma'}(\vec{k}+\vec{q})\rangle$ where $\ket{u_{\eta\sigma}(\vec{k})}$ are the Bloch wavefunctions of the active band of the continuum model \cite{Zhang_2010} and $\rho$ is a $4\times4$ matrix in spin and valley space. For simplicity, we assume the coupling matrices $\rho_j$ before band projection are momentum independent. For instance, ferromagnetic order-parameter fluctuations correspond to $\rho_j=(\sigma_x,\sigma_y,\sigma_z)\eta_0$. To also describe the attractive interactions coming from a phonon mode which transforms trivially under all symmetries, we will consider $\rho_j=\sigma_0\eta_0$.

\vspace{1em}
\textit{Candidate pairing states}---Before proceeding with energetics, let us classify the possible pairing instabilities based on symmetries. In its most general form, the superconducting order parameter $\Delta_{\vec{k},\vec{q}}$ is a matrix in spin and valley space and couples as
\begin{equation}
\mathcal{H}_{\text{MF}}=\sum_{\vec{k},\vec{q}}c^\dagger_{\vec{k}+\frac{\vec{q}}{2},\eta,\sigma}\left(\Delta_{\vec{k},\vec{q}}\right)_{\eta\sigma,\eta'\sigma'}c^\dagger_{-\vec{k}+\frac{\vec{q}}{2},\eta',\sigma'} +\text{H.c.} \label{MeanFieldSCHam}
\end{equation}
to the electrons of the active band in \equref{Eq:kineticterm}. Fermionic anti-symmetry requires $\Delta_{\vec{k},\vec{q}}=-\left(\Delta_{-\vec{k},\vec{q}}\right)^T$ and the presence of U(1)$_v$ symmetry implies that pairing is either entirely inter- or intravalley. 

First focusing on $\vec{q}=0$ in \equref{MeanFieldSCHam}, the associated Cooper pairs have vanishing or lattice-commensurate COM momentum $2\boldsymbol{K}$ for inter- or intravalley pairing, respectively. In both cases, $C_{3z}$ is preserved and we can use its irreducible representations (IRs) $A$, $E$, and $E^*$ to classify pairing. While the latter two are degenerate in the presence of time-reversal symmetry, this is not the case for $\mu_v \neq 0$. If $E$ is favored in one valley (one sign of $\mu_v$), time-reversal implies that $E^*$ is favored in the other (opposite sign of $\mu_v$). To make a meaningful statement, we compute the Chern number $C$ of the superconducting state, which can be expressed as a low-energy quantity in the single ``active’’ valley using that time-reversal symmetry is preserved before the spontaneous emergence of valley polarization in the normal state. We also compute the net Berry flux $\Phi_N$ within the continuum model, to provide a measure for its chirality. We then define the ``valley-independent Chern number'' as $\bar{C} := \text{sign} (\Phi_N) C$, which, as opposed to $C$ itself, is independent of the sign of $\mu_v$. We checked by acting with the microscopic representation of $C_{3z}$ on the superconducting states in our numerics that the $A$ state is achiral, $\bar{C}=C=0$, while $E$ ($E^*$) has $\bar{C}=1$ ($\bar{C}=-1$), see also \refcite{TheSTMPaper}. A positive (negative) $\bar{C}$ corresponds to a ferromagnetic (anti-ferromagnetic) coupling where the superconductor's chirality is the same as (opposite to) that of the normal state it emerges from. To obtain some general understanding of the expected sign of that coupling for an attractive interaction as mediated by phonons with $\rho=\mathbbm{1}$, we take a normal-state dispersion with $\epsilon_{\vec{k}} = \epsilon_{-\vec{k}}$, parametrize the superconducting order parameter by its angle $\varphi$ on the Fermi surface, to write the linearized gap equation as $\Delta_{\varphi} = \hat{K} \Delta_{\varphi}$. Expanding to second order in the momentum transfer of the interaction (App.~\ref{app:chirality}), we obtain $\hat{K} = \chi_0 - i(\mathcal{A}_\varphi + \mathcal{A}_{\varphi+\pi}) k_F \chi_2 \partial_{\varphi} + \chi_2 \partial_{\varphi}^2/2$ with $\chi_{0,2} > 0$ and $\mathcal{A}_\varphi$ denoting the Berry connection along the Fermi surface. We can thus see that the sign of the latter determines whether positive or negative $m$ are favored in $\Delta_\varphi \propto e^{im\varphi}$. Say $\mathcal{A}_\varphi >0$ (in a smooth gauge) such that $\Phi_N = \int\diff \varphi \mathcal{A}_\varphi >0$; from the form of $\hat{K}$, we conclude that $m>0$ will be favored, in turn leading to $\bar{C} = C > 0$ (ferromagnetic) or $\bar{C} < 0$ (antiferromagnetic) for an electron- (relevant to us here) and hole-like pocket, respectively.


As we will see below, where $\epsilon_{\vec{k}} \neq \epsilon_{-\vec{k}}$, also $\vec{q}\neq 0$ can be favored, corresponding to an incommensurate Cooper pair momentum. We will first start with a 1-$\vec{q}$ state, i.e., $\Delta_{\vec{k},\vec{q}} = \delta_{\vec{q},\vec{q}_0} \Delta_{\vec{k}}^{\vec{q}_0}$ with $\vec{q}_0 \neq 0$, which preserves translational symmetry in gauge-invariant observables. Importantly, at fixed $\vec{q}_0 \neq 0$, we cannot use $C_{3z}$ and its IRs to classify pairing but we can still use $\bar{C}$. 
Finally note that, for $\mu_s=0$, all states further decay into spin-singlet and spin-triplet.

\vspace{1em}
\textit{Leading 1-$\vec{q}$ Instabilities}---For a given choice of fluctuations as specified by $\rho_j$, we study the superconducting instabilities by solving the linearized mean-field gap equation of the normal-state Hamiltonian (\ref{Eq:kineticterm}) supplemented with the electron-electron interaction generated by integrating out the bosons in \equref{Eq:fbcoupling}. We choose a Lorentzian static susceptibility $\chi(\vec{q})$ for the bosons of the form $\chi(\vec{q})=g^2 \frac{\alpha}{\alpha^2+|\vec{q}|^2}$, where $\alpha$ can be thought of as a parameter quantifying how close the system is to criticality and $g^2$ represents the overall strength of the interaction relative to the kinetic energy. 

Once $\mu_v\neq0$ and time-reversal symmetry is broken, the states at $\vec{k}$ and $-\vec{k}$ are no longer degenerate and consequently it is no longer necessarily energetically favorable to pair electrons at opposite momenta, and non-zero $\vec{q}$ can become favorable. We note the precise value of $\vec{q}$ strongly depends on details of the Fermi surface; however, as illustrated in Fig.~\ref{fig:fig1}b, $\vec{q}$ is often approximately associated with the wave-vector for which there is the greatest approximate overlap between the shifted Fermi surfaces defined as $\epsilon_{\frac{\vec{q}}{2}\pm\vec{k},\eta,\sigma}=0$. To compute the value of $\vec{q}$ and the associated order parameter $\Delta_{\vec{k}}^{\vec{q}}$, we solve the linearized gap equation for different values of $\vec{q}$ and determine the leading instability from the value of $\vec{q}$ with the largest leading eigenvalue $\lambda$, see \figref{fig:fig1}c. A finite critical value of $\mu_v$ is necessary to obtain $\vec{q}\neq 0$, as a result of $C_{3z}$ symmetry \cite{scammell_theory_2022}, although as shown  in \figref{fig:fig1}d, this critical value is small and beyond our numerical resolution for the underlying normal state we study.

\begin{figure}
    \centering
    \includegraphics[width=\linewidth]{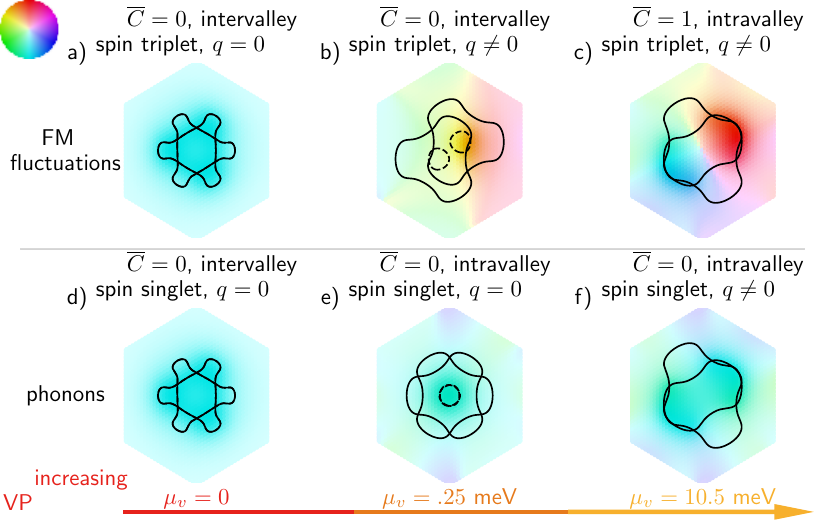}
    \caption{We show the highest eigenvalue solutions of the linearized gap equation for spin ferromagnetic fluctuations (a-c) and phonons (d-f) for varying degree of valley polarization $\mu_v$, with $\mu_s=0$. For each solution, we show a single, nonzero magnitude component of $\Delta_{\vec{k}}^{\vec{q}}$ in the projected space of the active band, where the intensity and color indicate the the magnitude and phase of $\Delta_{\vec{k}}^{\vec{q}}$ respectively. We also indicate $\overline{C}$ and the state's spin structure. We take a displacement field $D=110$ meV. For a discussion of how the gauge of the Bloch wavefunctions is fixed, see App.~\ref{App:LGE}. }
    \label{fig:fig2}
\end{figure}

While this behavior of $\vec{q}$ is qualitatively the same for all pairing mechanisms we considered, the $\vec{k}$ dependence of $\Delta_{\vec{k}}^{\vec{q}}$ is sensitive to the type of bosons mediating the interaction. Starting with ferromagnetic fluctuations, $\rho_j=(\sigma_x,\sigma_y,\sigma_z)\eta_0$, our results for varying degree of valley splitting $\mu_v$ are summarized in Fig.~\ref{fig:fig2}a-c. As expected for spin fluctuations, we find triplet pairing to be dominant.  In the intervalley pairing regime at weak $\mu_v$, the anti-symmetry $\Delta_{\vec{k}}^{\vec{q}}=-\left(\Delta_{-\vec{k}}^{\vec{q}}\right)^T$ does not constrain $\Delta_{\vec{k}}^{\vec{q}}$ to be even or odd in $\vec{k}$ (see \figref{fig:fig2}b) due to the valley degree of freedom. This changes for larger $\mu_v$, where intravalley pairing is energetically favored and, hence, the triplet order parameter must be odd in $\vec{k}$, as one can see in \figref{fig:fig2}c. The state in \figref{fig:fig2}a clearly transforms under IR $A$ but, as noted above, we cannot use the IRs of $C_{3z}$ to classify the pairing states at finite $\vec{q}$ as becomes apparent, e.g., in \figref{fig:fig2}b. Instead, we use the signed Chern number $\bar{C}$ of the superconductor and find that the initial $\bar{C}=0$ state transitions to a superconductor with $\bar{C}=1$, representing a ferromagnetic coupling between the normal state's and superconducting chirality.

The behavior is different for phonons, $\rho_j=\sigma_0\eta_0$. In this case, the leading instability also condenses at finite $\vec{q}$ above a critical value of $\mu_v$, but remains topologically trivial ($\overline{C}=0$) as valley polarization increases in Fig.~\ref{fig:fig2}d-f, with vanishing Chern number, $\overline{C}=0$ (see App.~\ref{app:Chern} for definition of $\bar{C}$ for the case with Fermi surfaces in both valleys).

\begin{figure}
    \centering
    \includegraphics[width=\linewidth]{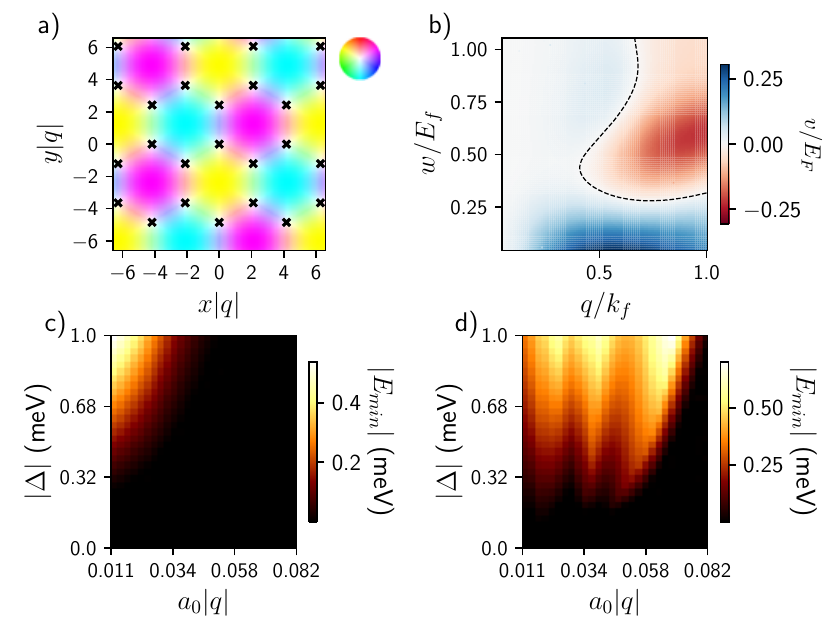}
    \caption{In a), we show the phase (color) and magnitude (intensity) in real space of a 3-$\vec{q}$ state order parameter $\Delta_{\vec{r},\vec{q}}$. In  b) we plot  $v/E_F$, relevant to the Landau free energy in Eq.~\eqref{eq: Landau th}, as a function of center of mass momentum $\vec{q}$ and $w$. We also compare the minimum excitation energy for the single-$\vec{q}$ (c) and three-$\vec{q}$ (d) for an intravalley spin-singlet gap function of the form $\Delta_{\vec{k}}=|\Delta|$. }
    \label{fig:fig3}
\end{figure}

\vspace{1em}
\textit{3-$\vec{q}$ states}---So far, we just assumed that only one $\vec{q}$ in \equref{MeanFieldSCHam} contributes. However, as can be seen in \figref{fig:fig1}, $C_{3z}$ symmetry requires that at least three $\vec{q}$ be degenerate in the $\vec{q}\neq 0$ regime. In our numerical calculations, we find exactly three degenerate maxima in most cases (see Fig.~\ref{fig:maxEig}). To take this into account, we write 
$\Delta_{\vec{k},\vec{q}} = \sum_{n=1}^3 \phi_n \delta_{\vec{q},\vec{q}_n} \Delta_{\vec{k}}^{\vec{q}_n}$, $\Delta_{\vec{k}}^{\vec{q}_n}$ being the eigenvector of the linearized gap equation,
$\vec{q}_n = C_{3z}^{n-1} \vec{q}_1$, and expand the free-energy as~\cite{LandauFootNote}
\begin{equation}\label{eq: Landau th}
    \mathcal{F}\sim r \sum_{n=1}^3 |\phi_n|^2 + u \left(\sum_{n=1}^3 |\phi_n|^2\right)^2 + v\sum_{n,m>n}|\phi_n|^2|\phi_m|^2
\end{equation}
where $u>\mathrm{min}\{0,-v\}$ as required by stability and we assumed a lattice-incommensurate value of $\vec{q}_n$. In the superconducting state, we have $r<0$ and the sign of $v$ determines whether we will have a 1-$\vec{q}$ ($v>0$) with only one of $\phi_{1,2,3}$ non-zero and the properties discussed above, or a 3-$\vec{q}$ ($v<0$) state, characterized by $|\phi_1|=|\phi_2|=|\phi_3|$. Note that the relative phase of $\phi_n$ just determines the origin of the emergent superconducting superlattice which preserves $C_{3z}$. As shown in Fig.~\ref{fig:fig3}(a), the 3-$\vec{q}$ state exhibits an emergent vortex-anti-vortex lattice in real space, which leads to complex spatially varying tunneling signatures \cite{TheSTMPaper}. The phases of theory~\eqref{eq: Landau th} can be characterized by three gauge invariant \textit{composite} order parameters $\varrho_{mn}=\phi_m\phi_n^*$ ($m\neq n$), describing charge density modulations with wave vector $\vec{q}_m-\vec{q}_n$~\cite{PDWReview,Zhou_2022}. It is obvious to see that the single-$\vec{q}$ phase has all $\varrho_{mn}=0$, while the 3-$\vec{q}$ state has all three $\varrho_{mn}$ nonzero and equal in absolute values. 

To understand how microscopic parameters influence the nature of the finite-$\vec{q}$ superconducting state, we consider a continuum model of spinless electrons with a $\vec{k}^2/(2m)$ dispersion coupled to a $p+ip$-wave superconducting order parameter which carries a finite COM momentum. Additionally, we deform the electron dispersion with a \textit{trigonal warping} term, whose strength we parametrize by $w$. 
Integrating out the electrons and expanding to quartic order in $\phi_n$, we get expressions for the coefficients of Eq.~\eqref{eq: Landau th} (App.~\ref{app: Landau theory}). 
In Fig.~\ref{fig:fig3}(b), we show $v$ as a function of $w$ and $|\vec{q}|$. We observe that at fixed $w$, a finite COM momentum is required to stabilize the 3-$\vec{q}$ state.

\vspace{1em}
\textit{Spectra}---Finally, we will discuss the superconducting spectra in more detail. Starting with $\vec{q}=0$, any infinitesimal order parameter with $|\Delta_{\vec{k}}^{\vec{q}=0}| \neq 0$ will immediately gap out the spectrum when $\mu_v=0$. However, when $\mu_v\neq0$---similar to the case of inter-band pairing in twisted bilayer graphene \cite{Christos_2023,putzer2025eliashbergtheorysuperfluidstiffness}, where the states forming Cooper pairs are not degenerate either since they are in different bands---a finite amount of pairing strength is required for a full gap; below that, the system will generically have Bogoliubov Fermi surfaces \cite{PhysRevB.98.224509}. At the same time, the thermal phase transition can become first order \cite{Christos_2023,putzer2025eliashbergtheorysuperfluidstiffness} and whether there is a nodal regime close to $T_c$ or not depends on microscopic details.
We next turn to the scenario where $\vec{q}\neq0$. In the case of the 1-$\vec{q}$ state, as illustrated in Fig.~\ref{fig:fig3}c, the gap, which emerges at finite pairing strength $|\Delta|$, decreases and ultimately vanishes with increasing $|\vec{q}|$, as the splitting between the paired energies increases. 

To compute the spectrum of the 3-$\vec{q}$ state, we have to take into account the broken translational symmetry and the associated superlattice reconstruction. This is done in analogy to the continuum model of twisted multilayer graphene \cite{dos2007graphene,bistritzer2011moire}, where the superlattice now comes from the pairing terms. In practice, we use a continuum model with 17 shells of $\vec{Q}$ vectors, which we find is sufficient to achieve convergence for our smallest $|\vec{q}|$ value. The resulting gap shown in Fig.~\ref{fig:fig3}b reveals that, similar to the case of the $1$-$\vec{q}$ state, a finite amount of pairing strength is needed at small $\vec{q}$ to generate a gap, with the threshold pairing strength increasing as $\vec{q}$ increases.

\vspace{1em}
\textit{Other normal states}---As at least one of the superconducting states in \refcite{Han2024a} is likely also spin polarized, we additionally comment on the effect of finite $\mu_s$ in \equref{Eq:kineticterm}. In our numerics, we found no qualitative changes in the features of the pairing states for $\mu_s \neq 0$ discussed so far, apart from sufficiently large $\mu_s$ leading to triplet pairing also for phonons. The main quantitative modification is that additional spin polarization lowers the critical $\mu_v$ for the formation of a topologically nontrivial superconductor ($\bar{C}=1$), see App.~\ref{app:OtherNormalState} and App.~\ref{AppH} for details. We also have studied the case of ferromagnetic fluctuations with a higher displacement field such that the fully spin and valley polarized normal state has an annular Fermi surface. In this case we find the dominant instability always has $C=0$, in agreement with the results of \cite{geier2024chiraltopologicalsuperconductivityisospin,yang2024topologicalincommensuratefuldeferrelllarkinovchinnikovsuperconductor,2025arXiv250217555Y,2025arXiv250305697M}. We also study fluctuations of a valley polarized order (see App.~\ref{App:VP}) with $\rho 
=\sigma_0\eta_z$ for varying $\mu_s$ and $\mu_v$. In this case we find the $A$ state dominates when $\mu_s=0$ with a finite $\mu_s$ leading to a transition to the $E$ state.

\vspace{1em}
\textit{Conclusion}---We analyzed superconducting instabilities within a continuum model description of rhombohedral tetra-layer graphene as a function of spin and valley polarization. The analysis reveals rich superconducting physics, including the possibility of topological superconductivity and the spontaneous emergence of a superconducting superlattice.  
As discussed in \cite{TheSTMPaper}, future scanning tunneling spectroscopy experiments will allow to probe key features of the identified candidate pairing states. We also note that a valley-imbalanced normal state will generically give rise to non-reciprocal superconductivity \cite{scammell_theory_2022}, with asymmetric current-voltage characteristics, providing a complementary way of accessing the reduced symmetries of the system. Finally, although our analysis explored only superconducting states, it would be interesting to consider other instabilities which may compete with superconductivity in the system.

\textit{Note added---}Recently, a preprint \cite{2025arXiv250305697M} appeared that also discusses the relative chirality between the normal and superconducting state. The results agree where there is overlap (phonon-mediated pairing).

\begin{acknowledgments}
The authors thank Subir Sachdev for comments on the manuscript. M.S.S.~further thanks Denis Sedov for a collaboration on a related project \cite{TheSTMPaper}.
M.C. acknowledges funding from NSF grant DMR-2245246. P.M.B. acknowledges support by the German National Academy of Sciences Leopoldina through Grant No.~LPDS 2023-06. This research is funded in part by the Gordon
and Betty Moore Foundation’s EPiQS Initiative, Grant
GBMF8683 to P.M.B. M.S.S.~further acknowledges funding by the European Union (ERC-2021-STG, Project 101040651---SuperCorr). Views and opinions expressed are however those of the authors only and do not necessarily reflect those of the European Union or the European Research Council Executive Agency. Neither the European Union nor the granting authority can be held responsible for them. M.S.S.~is also greatful for support by grant NSF PHY-2309135 for his stay at the Kavli Institute for Theoretical Physics (KITP) where a part of the research was done.
\end{acknowledgments}

\bibliography{draft_Refs}

@misc{Han2024a,
      title={Signatures of Chiral Superconductivity in Rhombohedral Graphene}, 
      author={Tonghang Han and Zhengguang Lu and Yuxuan Yao and Lihan Shi and Jixiang Yang and Junseok Seo and Shenyong Ye and Zhenghan Wu and Muyang Zhou and Haoyang Liu and Gang Shi and Zhenqi Hua and Kenji Watanabe and Takashi Taniguchi and Peng Xiong and Liang Fu and Long Ju},
      year={2024},
      eprint={2408.15233},
      archivePrefix={arXiv}
}

@misc{qin2024chiralfinitemomentumsuperconductivitytetralayer,
      title={Chiral finite-momentum superconductivity in the tetralayer graphene}, 
      author={Qiong Qin and Congjun Wu},
      year={2024},
      eprint={2412.07145},
      archivePrefix={arXiv},
      primaryClass={cond-mat.supr-con},
      url={https://arxiv.org/abs/2412.07145}, 
}

@misc{sau2024theoryanomaloushalleffect,
      title={Theory of anomalous Hall effect from screened vortex charge in a phase disordered superconductor}, 
      author={Jay D. Sau and Shuyang Wang},
      year={2024},
      eprint={2411.08969},
      archivePrefix={arXiv},
      primaryClass={cond-mat.supr-con},
      url={https://arxiv.org/abs/2411.08969}, 
}

@misc{yang2024topologicalincommensuratefuldeferrelllarkinovchinnikovsuperconductor,
      title={Topological incommensurate Fulde-Ferrell-Larkin-Ovchinnikov superconductor and Bogoliubov Fermi surface in rhombohedral tetra-layer graphene}, 
      author={Hui Yang and Ya-Hui Zhang},
      year={2024},
      eprint={2411.02503},
      archivePrefix={arXiv},
      primaryClass={cond-mat.supr-con},
      url={https://arxiv.org/abs/2411.02503}, 
}

@misc{geier2024chiraltopologicalsuperconductivityisospin,
      title={Chiral and topological superconductivity in isospin polarized multilayer graphene}, 
      author={Max Geier and Margarita Davydova and Liang Fu},
      year={2024},
      eprint={2409.13829},
      archivePrefix={arXiv},
      primaryClass={cond-mat.supr-con},
      url={https://arxiv.org/abs/2409.13829}, 
}

@article{Kim_2025,
   title={Topological chiral superconductivity beyond pairing in a Fermi liquid},
   volume={111},
   ISSN={2469-9969},
   url={http://dx.doi.org/10.1103/PhysRevB.111.014508},
   DOI={10.1103/physrevb.111.014508},
   number={1},
   journal={Physical Review B},
   publisher={American Physical Society (APS)},
   author={Kim, Minho and Timmel, Abigail and Ju, Long and Wen, Xiao-Gang},
   year={2025},
   month=jan }

@misc{chou2024intravalleyspinpolarizedsuperconductivityrhombohedral,
      title={Intravalley spin-polarized superconductivity in rhombohedral tetralayer graphene}, 
      author={Yang-Zhi Chou and Jihang Zhu and Sankar Das Sarma},
      year={2024},
      eprint={2409.06701},
      archivePrefix={arXiv},
      primaryClass={cond-mat.supr-con},
      url={https://arxiv.org/abs/2409.06701}, 
}

@article{Ghazaryan_2023,
   title={Multilayer graphenes as a platform for interaction-driven physics and topological superconductivity},
   volume={107},
   ISSN={2469-9969},
   url={http://dx.doi.org/10.1103/PhysRevB.107.104502},
   DOI={10.1103/physrevb.107.104502},
   number={10},
   journal={Physical Review B},
   publisher={American Physical Society (APS)},
   author={Ghazaryan, Areg and Holder, Tobias and Berg, Erez and Serbyn, Maksym},
   year={2023},
   month=mar }

@article{Zhang_2010,
  title = {Band structure of $ABC$-stacked graphene trilayers},
  author = {Zhang, Fan and Sahu, Bhagawan and Min, Hongki and MacDonald, A. H.},
  journal = {Phys. Rev. B},
  volume = {82},
  issue = {3},
  pages = {035409},
  numpages = {10},
  year = {2010},
  month = {Jul},
  publisher = {American Physical Society},
  doi = {10.1103/PhysRevB.82.035409},
  url = {https://link.aps.org/doi/10.1103/PhysRevB.82.035409}
}

@article{Alicea_2012,
doi = {10.1088/0034-4885/75/7/076501},
url = {https://dx.doi.org/10.1088/0034-4885/75/7/076501},
year = {2012},
month = {jun},
publisher = {IOP Publishing},
volume = {75},
number = {7},
pages = {076501},
author = {Alicea, Jason},
title = {New directions in the pursuit of Majorana fermions in solid state systems},
journal = {Reports on Progress in Physics}
}

@article{PhysRevB.80.165409,
  title = {Trigonal warping and Berry's phase $N\ensuremath{\pi}$ in ABC-stacked multilayer graphene},
  author = {Koshino, Mikito and McCann, Edward},
  journal = {Phys. Rev. B},
  volume = {80},
  issue = {16},
  pages = {165409},
  numpages = {8},
  year = {2009},
  month = {Oct},
  publisher = {American Physical Society},
  doi = {10.1103/PhysRevB.80.165409},
  url = {https://link.aps.org/doi/10.1103/PhysRevB.80.165409}
}

@article{dos2007graphene,
  title={Graphene bilayer with a twist: electronic structure},
  author={Dos Santos, J M B Lopes and Peres, N M R and Neto, A H Castro},
  journal={Phys. Rev. Lett.},
  volume={99},
  number={25},
  pages={256802},
  year={2007},
  publisher={APS},
  doi={10.1103/PhysRevLett.99.256802}
}

@article{bistritzer2011moire,
  title={Moir{\'e} bands in twisted double-layer graphene},
  author={Bistritzer, Rafi and MacDonald, Allan H},
  journal={Proc. Natl. Acad. Sci. U.S.A.},
  volume={108},
  number={30},
  pages={12233--12237},
  year={2011},
  publisher={National Acad Sciences},
  doi={10.1073/pnas.1108174108}
}

@article{review1,
	author = {Andrei, Eva Y. and MacDonald, Allan H.},
	date = {2020/12/01},
	doi = {10.1038/s41563-020-00840-0},
	id = {Andrei2020},
	isbn = {1476-4660},
	journal = {Nature Materials},
	number = {12},
	pages = {1265--1275},
	title = {Graphene bilayers with a twist},
	url = {https://doi.org/10.1038/s41563-020-00840-0},
	volume = {19},
	year = {2020}}

@article{review2,
	author = {Balents, Leon and Dean, Cory R. and Efetov, Dmitri K. and Young, Andrea F.},
	date = {2020/07/01},
	doi = {10.1038/s41567-020-0906-9},
	id = {Balents2020},
	isbn = {1745-2481},
	journal = {Nature Physics},
	number = {7},
	pages = {725--733},
	title = {Superconductivity and strong correlations in moir{\'e} flat bands},
	url = {https://doi.org/10.1038/s41567-020-0906-9},
	volume = {16},
	year = {2020}}

@article{review3,
	author = {Nuckolls, Kevin P. and Yazdani, Ali},
	date = {2024/07/01},
	doi = {10.1038/s41578-024-00682-1},
	id = {Nuckolls2024},
	isbn = {2058-8437},
	journal = {Nature Reviews Materials},
	number = {7},
	pages = {460--480},
	title = {A microscopic perspective on moir{\'e}materials},
	url = {https://doi.org/10.1038/s41578-024-00682-1},
	volume = {9},
	year = {2024}}

@ARTICLE{2025arXiv250309692B,
       author = {{Bernevig}, B. Andrei and {Kwan}, Yves H.},
        title = "{``Berry Trashcan'' Model of Interacting Electrons in Rhombohedral Graphene}",
      journal = {arXiv e-prints},
         year = 2025,
        month = mar,
archivePrefix = {arXiv},
       eprint = {2503.09692},
 primaryClass = {cond-mat.str-el},
       adsurl = {https://ui.adsabs.harvard.edu/abs/2025arXiv250309692B}
}

@article{PDWReview,
   author = "Agterberg, Daniel F. and Davis, J.C. Séamus and Edkins, Stephen D. and Fradkin, Eduardo and Van Harlingen, Dale J. and Kivelson, Steven A. and Lee, Patrick A. and Radzihovsky, Leo and Tranquada, John M. and Wang, Yuxuan",
   title = "The Physics of Pair-Density Waves: Cuprate Superconductors and Beyond", 
   journal= "Annual Review of Condensed Matter Physics",
   year = "2020",
   volume = "11",
   number = "Volume 11, 2020",
   pages = "231-270",
   doi = "https://doi.org/10.1146/annurev-conmatphys-031119-050711",
   url = "https://www.annualreviews.org/content/journals/10.1146/annurev-conmatphys-031119-050711",
   publisher = "Annual Reviews",
   issn = "1947-5462",
   type = "Journal Article"
  }

@ARTICLE{TheSTMPaper,
       author = {{Sedov}, Denis and {Scheurer}, Mathias S.},
        title = "{Probing superconductivity with tunneling spectroscopy in rhombohedral graphene}",
      journal = {arXiv e-prints},
         year = 2024,
        month = jul,
archivePrefix = {arXiv},
       eprint = {2503.12650},
 primaryClass = {cond-mat.supr-con}
}

@article{PhysRevB.105.L081407,
  title = {Metals, fractional metals, and superconductivity in rhombohedral trilayer graphene},
  author = {Szab\'o, Andr\'as L. and Roy, Bitan},
  journal = {Phys. Rev. B},
  volume = {105},
  issue = {8},
  pages = {L081407},
  numpages = {5},
  year = {2022},
  month = {Feb},
  publisher = {American Physical Society},
  doi = {10.1103/PhysRevB.105.L081407},
  url = {https://link.aps.org/doi/10.1103/PhysRevB.105.L081407}
}

@ARTICLE{2025arXiv250305697M,
       author = {{May-Mann}, Julian and {Helbig}, Tobias and {Devakul}, Trithep},
        title = "{How pairing mechanism dictates topology in valley-polarized superconductors with Berry curvature}",
      journal = {arXiv e-prints},
         year = 2025,
        month = mar,
       eprint = {2503.05697},
 primaryClass = {cond-mat.supr-con},
       adsurl = {https://ui.adsabs.harvard.edu/abs/2025arXiv250305697M}
}

@ARTICLE{2024arXiv241109664J,
       author = {{Jahin}, Ammar and {Lin}, Shi-Zeng},
        title = "{Enhanced Kohn-Luttinger topological superconductivity in bands with nontrivial geometry}",
      journal = {arXiv e-prints},
         year = 2024,
        month = nov,
archivePrefix = {arXiv},
       eprint = {2411.09664},
 primaryClass = {cond-mat.supr-con},
       adsurl = {https://ui.adsabs.harvard.edu/abs/2024arXiv241109664J}
}

@article{Christos_2023,
   title={Nodal band-off-diagonal superconductivity in twisted graphene superlattices},
   volume={14},
   ISSN={2041-1723},
   url={http://dx.doi.org/10.1038/s41467-023-42471-4},
   DOI={10.1038/s41467-023-42471-4},
   number={1},
   journal={Nature Communications},
   publisher={Springer Science and Business Media LLC},
   author={Christos, Maine and Sachdev, Subir and Scheurer, Mathias S.},
   year={2023},
   month=nov }

@misc{putzer2025eliashbergtheorysuperfluidstiffness,
      title={Eliashberg Theory and Superfluid Stiffness of Band-Off-Diagonal Pairing in Twisted Graphene}, 
      author={Bernhard Putzer and Mathias S. Scheurer},
      year={2025},
      eprint={2501.12435},
      archivePrefix={arXiv},
      primaryClass={cond-mat.supr-con},
      url={https://arxiv.org/abs/2501.12435}, 
}

@article{PhysRevB.98.224509,
  title = {Bogoliubov Fermi surfaces: General theory, magnetic order, and topology},
  author = {Brydon, P. M. R. and Agterberg, D. F. and Menke, Henri and Timm, C.},
  journal = {Phys. Rev. B},
  volume = {98},
  issue = {22},
  pages = {224509},
  numpages = {24},
  year = {2018},
  month = {Dec},
  publisher = {American Physical Society},
  doi = {10.1103/PhysRevB.98.224509},
  url = {https://link.aps.org/doi/10.1103/PhysRevB.98.224509}
}

@article{Zhou_2022,
   title={Chern Fermi pocket, topological pair density wave, and charge-4e and charge-6e superconductivity in kagomé superconductors},
   volume={13},
   ISSN={2041-1723},
   url={http://dx.doi.org/10.1038/s41467-022-34832-2},
   DOI={10.1038/s41467-022-34832-2},
   number={1},
   journal={Nature Communications},
   publisher={Springer Science and Business Media LLC},
   author={Zhou, Sen and Wang, Ziqiang},
   year={2022},
   month=nov }

@article{scammell_theory_2022,
  title = {Theory of Zero-Field Superconducting Diode Effect in Twisted Trilayer Graphene},
  author = {Scammell, Harley D. and Li, J. I. A. and Scheurer, Mathias S.},
  year = {2022},
  month = mar,
  journal = {2D Materials},
  volume = {9},
  number = {2},
  pages = {025027},
  publisher = {{IOP Publishing}},
  issn = {2053-1583},
  doi = {10.1088/2053-1583/ac5b16},
  urldate = {2023-02-11},
  langid = {english},
}

@article{ShubhayuPairing,
	author = {Chatterjee, Shubhayu and Wang, Taige and Berg, Erez and Zaletel, Michael P.},
	date = {2022/10/12},
	doi = {10.1038/s41467-022-33561-w},
	id = {Chatterjee2022},
	isbn = {2041-1723},
	journal = {Nature Communications},
	number = {1},
	pages = {6013},
	title = {Inter-valley coherent order and isospin fluctuation mediated superconductivity in rhombohedral trilayer graphene},
	url = {https://doi.org/10.1038/s41467-022-33561-w},
	volume = {13},
	year = {2022}}

@article{PhysRevLett.127.187001,
  title = {Acoustic-Phonon-Mediated Superconductivity in Rhombohedral Trilayer Graphene},
  author = {Chou, Yang-Zhi and Wu, Fengcheng and Sau, Jay D. and Das Sarma, Sankar},
  journal = {Phys. Rev. Lett.},
  volume = {127},
  issue = {18},
  pages = {187001},
  numpages = {6},
  year = {2021},
  month = {Oct},
  publisher = {American Physical Society},
  doi = {10.1103/PhysRevLett.127.187001},
  url = {https://link.aps.org/doi/10.1103/PhysRevLett.127.187001}
}

@article{PhysRevLett.127.247001,
  title = {Unconventional Superconductivity in Systems with Annular Fermi Surfaces: Application to Rhombohedral Trilayer Graphene},
  author = {Ghazaryan, Areg and Holder, Tobias and Serbyn, Maksym and Berg, Erez},
  journal = {Phys. Rev. Lett.},
  volume = {127},
  issue = {24},
  pages = {247001},
  numpages = {7},
  year = {2021},
  month = {Dec},
  publisher = {American Physical Society},
  doi = {10.1103/PhysRevLett.127.247001},
  url = {https://link.aps.org/doi/10.1103/PhysRevLett.127.247001}
}

@article{PhysRevB.105.134524,
  title = {Kohn-Luttinger superconductivity and intervalley coherence in rhombohedral trilayer graphene},
  author = {You, Yi-Zhuang and Vishwanath, Ashvin},
  journal = {Phys. Rev. B},
  volume = {105},
  issue = {13},
  pages = {134524},
  numpages = {12},
  year = {2022},
  month = {Apr},
  publisher = {American Physical Society},
  doi = {10.1103/PhysRevB.105.134524},
  url = {https://link.aps.org/doi/10.1103/PhysRevB.105.134524}
}

@article{PhysRevB.107.104502,
  title = {Multilayer graphenes as a platform for interaction-driven physics and topological superconductivity},
  author = {Ghazaryan, Areg and Holder, Tobias and Berg, Erez and Serbyn, Maksym},
  journal = {Phys. Rev. B},
  volume = {107},
  issue = {10},
  pages = {104502},
  numpages = {16},
  year = {2023},
  month = {Mar},
  publisher = {American Physical Society},
  doi = {10.1103/PhysRevB.107.104502},
  url = {https://link.aps.org/doi/10.1103/PhysRevB.107.104502}
}

@article{PhysRevB.107.L161106,
  title = {Superconductivity from electronic interactions and spin-orbit enhancement in bilayer and trilayer graphene},
  author = {Jimeno-Pozo, Alejandro and Sainz-Cruz, H\'ector and Cea, Tommaso and Pantale\'on, Pierre A. and Guinea, Francisco},
  journal = {Phys. Rev. B},
  volume = {107},
  issue = {16},
  pages = {L161106},
  numpages = {7},
  year = {2023},
  month = {Apr},
  publisher = {American Physical Society},
  doi = {10.1103/PhysRevB.107.L161106},
  url = {https://link.aps.org/doi/10.1103/PhysRevB.107.L161106}
}

@article{PhysRevLett.130.146001,
  title = {Functional Renormalization Group Study of Superconductivity in Rhombohedral Trilayer Graphene},
  author = {Qin, Wei and Huang, Chunli and Wolf, Tobias and Wei, Nemin and Blinov, Igor and MacDonald, Allan H.},
  journal = {Phys. Rev. Lett.},
  volume = {130},
  issue = {14},
  pages = {146001},
  numpages = {7},
  year = {2023},
  month = {Apr},
  publisher = {American Physical Society},
  doi = {10.1103/PhysRevLett.130.146001},
  url = {https://link.aps.org/doi/10.1103/PhysRevLett.130.146001}
}

@article{TwistedNonTwisted,
	author = {Pantale{\'o}n, Pierre A. and Jimeno-Pozo, Alejandro and Sainz-Cruz, H{\'e}ctor and Phong, V{\~o}Tiến and Cea, Tommaso and Guinea, Francisco},
	date = {2023/05/01},
	doi = {10.1038/s42254-023-00575-2},
	id = {Pantale{\'o}n2023},
	isbn = {2522-5820},
	journal = {Nature Reviews Physics},
	number = {5},
	pages = {304--315},
	title = {Superconductivity and correlated phases in non-twisted bilayer and trilayer graphene},
	url = {https://doi.org/10.1038/s42254-023-00575-2},
	volume = {5},
	year = {2023}}

@article{PhysRevB.108.045404,
  title = {Charge fluctuations, phonons, and superconductivity in multilayer graphene},
  author = {Li, Ziyan and Kuang, Xueheng and Jimeno-Pozo, Alejandro and Sainz-Cruz, H\'ector and Zhan, Zhen and Yuan, Shengjun and Guinea, Francisco},
  journal = {Phys. Rev. B},
  volume = {108},
  issue = {4},
  pages = {045404},
  numpages = {11},
  year = {2023},
  month = {Jul},
  publisher = {American Physical Society},
  doi = {10.1103/PhysRevB.108.045404},
  url = {https://link.aps.org/doi/10.1103/PhysRevB.108.045404}
}

@article{PhysRevB.108.134503,
  title = {Superconductivity near spin and valley orders in graphene multilayers},
  author = {Dong, Zhiyu and Levitov, Leonid and Chubukov, Andrey V.},
  journal = {Phys. Rev. B},
  volume = {108},
  issue = {13},
  pages = {134503},
  numpages = {12},
  year = {2023},
  month = {Oct},
  publisher = {American Physical Society},
  doi = {10.1103/PhysRevB.108.134503},
  url = {https://link.aps.org/doi/10.1103/PhysRevB.108.134503}
}

@ARTICLE{2024arXiv240617036D,
       author = {{Dong}, Zhiyu and {Lantagne-Hurtubise}, {\'E}tienne and {Alicea}, Jason},
        title = "{Superconductivity from spin-canting fluctuations in rhombohedral graphene}",
      journal = {arXiv e-prints},
         year = 2024,
        month = jun,
archivePrefix = {arXiv},
       eprint = {2406.17036},
 primaryClass = {cond-mat.supr-con},
       adsurl = {https://ui.adsabs.harvard.edu/abs/2024arXiv240617036D}
}

@article{SCTrilayer,
	author = {Zhou, Haoxin and Xie, Tian and Taniguchi, Takashi and Watanabe, Kenji and Young, Andrea F.},
	date = {2021/10/01},
	doi = {10.1038/s41586-021-03926-0},
	id = {Zhou2021},
	isbn = {1476-4687},
	journal = {Nature},
	number = {7881},
	pages = {434--438},
	title = {Superconductivity in rhombohedral trilayer graphene},
	url = {https://doi.org/10.1038/s41586-021-03926-0},
	volume = {598},
	year = {2021}}

@article{HalfQuarterMetals,
	author = {Zhou, Haoxin and Xie, Tian and Ghazaryan, Areg and Holder, Tobias and Ehrets, James R. and Spanton, Eric M. and Taniguchi, Takashi and Watanabe, Kenji and Berg, Erez and Serbyn, Maksym and Young, Andrea F.},
	date = {2021/10/01},
	doi = {10.1038/s41586-021-03938-w},
	id = {Zhou2021},
	isbn = {1476-4687},
	journal = {Nature},
	number = {7881},
	pages = {429--433},
	title = {Half- and quarter-metals in rhombohedral trilayer graphene},
	url = {https://doi.org/10.1038/s41586-021-03938-w},
	volume = {598},
	year = {2021}}

@article{PhysRevResearch.6.043240,
  title = {Ideal Chern bands with strong short-range repulsion: Applications to correlated metals, superconductivity, and topological order},
  author = {Wilhelm, Patrick H. and L\"auchli, Andreas M. and Scheurer, Mathias S.},
  journal = {Phys. Rev. Res.},
  volume = {6},
  issue = {4},
  pages = {043240},
  numpages = {20},
  year = {2024},
  month = {Dec},
  publisher = {American Physical Society},
  doi = {10.1103/PhysRevResearch.6.043240},
  url = {https://link.aps.org/doi/10.1103/PhysRevResearch.6.043240}
}

@ARTICLE{2025arXiv250219474P,
       author = {{Parra-Martinez}, Guillermo and {Jimeno-Pozo}, Alejandro and {Tien Phong}, Vo and {Sainz-Cruz}, Hector and {Kaplan}, Daniel and {Emanuel}, Peleg and {Oreg}, Yuval and {Pantaleon}, Pierre A. and {Silva-Guillen}, Jose Angel and {Guinea}, Francisco},
        title = "{Band Renormalization, Quarter Metals, and Chiral Superconductivity in Rhombohedral Tetralayer Graphene}",
      journal = {arXiv e-prints},
         year = 2025,
        month = feb,
       eprint = {2502.19474},
 primaryClass = {cond-mat.str-el},
       adsurl = {https://ui.adsabs.harvard.edu/abs/2025arXiv250219474P}
}

@ARTICLE{2025arXiv250217555Y,
       author = {{Yoon}, Chiho and {Xu}, Tianyi and {Barlas}, Yafis and {Zhang}, Fan},
        title = "{Quarter Metal Superconductivity}",
      journal = {arXiv e-prints},
         year = 2025,
        month = feb,
archivePrefix = {arXiv},
       eprint = {2502.17555},
 primaryClass = {cond-mat.mes-hall},
       adsurl = {https://ui.adsabs.harvard.edu/abs/2025arXiv250217555Y}
}

@misc{LandauFootNote,
note={Note that in the spin unpolarized case, due to the system's SU(2) invariance under spin rotations, one can write the order parameter as $\Delta_{\boldsymbol{k},\boldsymbol{q}}^{\eta\sigma,\eta'\sigma'}$ as $\sum_{n=1}^3 \myvect{\phi}_n\cdot(i\sigma_y\myvect{\sigma})_{\sigma\sigma'}\delta_{\eta\eta'}\Delta_{\boldsymbol{k}}^{\boldsymbol{q}_n}\,\delta_{\vec{q},\vec{q}_n}$, thereby promoting $\myvect{\phi}_n$ to a complex \textit{vector}. In this case the Landau theory is invariant under O(3) rotations of $\myvect{\phi}_n$ and it allows for more quartic terms than the ones in Eq.~\eqref{eq: Landau th}. Additionally, more phases than the two discussed here are in principle possible.}
}

@article{ZHan2022,
  title = {Pair density wave and reentrant superconducting tendencies originating from valley polarization},
  author = {Han, Zhaoyu and Kivelson, Steven A.},
  journal = {Phys. Rev. B},
  volume = {105},
  issue = {10},
  pages = {L100509},
  numpages = {5},
  year = {2022},
  month = {Mar},
  publisher = {American Physical Society},
  doi = {10.1103/PhysRevB.105.L100509},
  url = {https://link.aps.org/doi/10.1103/PhysRevB.105.L100509}
}

@misc{abanov2001quantumcriticaltheoryspinfermionmodel,
      title={Quantum-critical theory of the spin-fermion model and its application to cuprates. Normal state analysis}, 
      author={Ar. Abanov and Andrey V. Chubukov and J. Schmalian},
      year={2001},
      eprint={cond-mat/0107421},
      archivePrefix={arXiv},
      primaryClass={cond-mat.supr-con},
      url={https://arxiv.org/abs/cond-mat/0107421}, 
}

\onecolumngrid

\begin{appendix}

\section{Model and Linearized Gap Equation}\label{App:LGE}
In this appendix, we describe the microscopic continuum model we take as our starting point and give the full gap equation we solve in the main text. We take the model of \cite{Zhang_2010,Ghazaryan_2023} in a single valley:
\begin{equation}\label{Eq:continuumModel}
    \mathcal{H}=\sum_{\vec{k}}c_{\vec{k}}^\dagger\mathcal{H}_{\vec{k}}c_{\vec{k}} \qquad \mathcal{H}_{\vec{k}}=\begin{pmatrix}
        \delta+D/2 & v_0\pi^\dagger & v_4 \pi^\dagger & v_3\pi & 0 & \gamma_2/2 &0 & 0\\v_0\pi & D/2 & \gamma_1 & v_4\pi^\dagger & 0 & 0 & 0 & 0\\ v_4 \pi & \gamma_1 & D/6 & v_0\pi^\dagger & v_4\pi^\dagger & v_3\pi & 0 &\gamma_2/2\\v_3\pi^\dagger & v_4\pi & v_0\pi& D/6 & \gamma_1 & v_4\pi^\dagger & 0 & 0\\ 0 & 0 & v_4\pi & \gamma_1 & -D/6 & v_0\pi^\dagger & v_4\pi^\dagger & v_3\pi \\\gamma_2/2 & 0 & v_3\pi^\dagger & v_4\pi & v_0\pi & -D/6 & \gamma_1 & v_4\pi^\dagger\\0 & 0 & 0 & 0 & v_4\pi & \gamma_1 & -D/2 & v_0\pi^\dagger \\ 0 & 0 & \gamma_2/2 & 0 & v_3\pi^\dagger & v_4\pi & v_0\pi & \delta-D/2
    \end{pmatrix}
\end{equation}
where $\pi=k_x+ik_y$ and $v_i=\frac{\sqrt{3}a_0}{2}\gamma_i$ where $a_0=0.246$ nm. We use the parameter values of \cite{PhysRevB.107.104502}, taking $\gamma_0=3.1$ eV, $\gamma_1=0.38$ eV, $\gamma_2=-0.015$ eV, $\gamma_3=0.29$ eV, $\gamma_4=-0.141$ eV, and $\delta=-0.0105$ eV. The Hamiltonian in the opposite valley is obtained by acting a spinless time-reversal symmetry $\mathcal{T}$ on Eq.~\ref{Eq:continuumModel}, which acts on the electrons in the microscopic basis as $\mathcal{T}:c_{\vec{k},l,\rho,\eta}\rightarrow c^\dagger_{-\vec{k},l,\rho,-\eta}$ (here $\rho$ and $l=1,2,3,4$ label the sublattice and layer degrees of freedom respectively). There is also a threefold rotation symmetry we denote as $C_{3z}$ (which is always present, including when $\mu_v\neq0$) which acts as $C_{3z}:c_{\vec{k},\rho\eta}\rightarrow [e^{-i2\pi/3\left(\rho_z+l+1\right)}]_{\rho\eta;\rho'\eta'}\,c_{C_{3z}(\vec{k}),\rho'\eta'}$.

We then consider taking only the active band of Eq.~\ref{Eq:continuumModel} at small doping and add interactions generated by coupling the electrons to an arbitrary bosonic mode $\phi_j$ as in Eq.~\ref{Eq:fbcoupling}:
\begin{equation}
    \mathcal{H}_{int}=\frac{1}{A}\sum_{\vec{p}}\chi_{\vec{p}}\rho_j(\vec{p})\rho_j(-\vec{p})
\end{equation}
We decouple this interaction in the cooper channel allowing for nonzero expectation values of the bilinear (written in terms of the projected degrees of freedom $c_{\vec{k},\eta,\sigma}$ now labeled only by spin $\sigma$ and valley $\eta$):
\begin{equation}
    \langle c_{\vec{k},\eta\sigma}c_{\vec{k'},\eta'\sigma'}\rangle\sim\delta_{\vec{k}-\frac{\vec{q}}{2},-\vec{k'}+\frac{\vec{q}}{2}}
\end{equation}
In this channel, the mean-field Hamiltonian is:
\begin{equation}\label{Eq:decouple}
\mathcal{H}_{MF}=\mathcal{H}_{kin}+\sum_{\vec{k}}c^\dagger_{\vec{k}+\frac{\vec{q}}{2},\eta,\sigma}\left(\Delta_{\vec{k}}^{\vec{q}}\right)^{\eta\sigma,\eta'\sigma'}c_{-\vec{k}+\frac{\vec{q}}{2},\eta',\sigma'}+\text{h.c.}
\end{equation}
where $\mathcal{H}_{kin.}$ is the dispersion of the active band of Eq.~\ref{Eq:continuumModel} and as in the main text we have defined:
\begin{equation}\label{Eq:deltaDef}
    \left(\Delta_{\vec{k}}^{\vec{q}}\right)^{\eta\sigma,\eta'\sigma'}=\sum_{\vec{p}}\chi_{\vec{p}}\langle c_{-\vec{k}-\vec{p}+\frac{\vec{q}}{2},\tau\alpha}c_{\vec{k}+\vec{p}+\frac{\vec{q}}{2},\tau'\alpha'}\rangle F_{\vec{k}+\frac{\vec{q}}{2},\vec{p}}^{\eta\sigma,\tau'\alpha'}F_{-\vec{k}+\frac{\vec{q}}{2},-\vec{p}}^{\eta'\sigma'\tau\alpha}
\end{equation}
In our numerics, we fix the gauge of the bloch wavefunctions used to compute $F_{\vec{k}+\frac{\vec{q}}{2},\vec{p}}^{\eta\sigma,\tau'\alpha'}$ by choosing a test wave function of the continuum model $|u(\vec{k}^0)\rangle$ at momentum $|\vec{k}^0|\gg k_F$ and fixing the phases of the bloch wavefunctions such that $\langle u(\vec{k}^0)|u_{\eta}(\vec{k})\rangle/|\langle u(\vec{k}^0)|u_{\eta\sigma}(\vec{k})\rangle|=1$ for $\vec{k}$ within the extent we study numerically and wavefunctions $|u_{\eta\sigma}(\vec{k})\rangle$ in the active band of the continuum model at small doping. For intervalley pairings, we further use time-reversal symmetry to fix $\langle u_{+,\sigma}(\vec{k}^0)|u_{-,\sigma}(-\vec{k})^*\rangle=1$. We note that a variational mean-field solution for the gap $\left(\Delta_{\vec{k}}^{\vec{q}}\right)^{\eta\sigma,\eta'\sigma'}$ generically depends on two momenta, $\vec{k}$ and the COM momenta $\vec{q}$. Minimization of the free energy associated with Eq.~\ref{Eq:decouple} leads to the following self-consistency equation valid when $\Delta_{\vec{k}}^{\vec{q}}\rightarrow 0$:
\begin{equation}
\begin{split}
    \left(\Delta^{\vec{q}}_{\vec{k}}\right)^{\eta\sigma,\eta'\sigma'}&=-\frac{1}{A}\sum_{\vec{p}}\sum_{\tau,\tau'=\uparrow\downarrow,\alpha,\beta=\pm}\chi_{\vec{p}}F_{\vec{k}+\frac{\vec{q}}{2},\vec{p}}^{\eta\sigma,\tau'\beta}F_{-\vec{k}+\frac{\vec{q}}{2},-\vec{p}}^{\eta'\sigma',\tau\alpha}\left(\frac{n_F(\xi^{\tau'\beta}_{\vec{k}+\vec{p}+\frac{\vec{q}}{2}})-n_F(-\xi^{\tau\alpha}_{-\vec{k}-\vec{p}+\frac{\vec{q}}{2}})}{\xi^{\tau'\beta}_{\vec{k}+\vec{q}+\frac{\vec{q}}{2}}+\xi^{\tau\alpha}_{-\vec{k}-\vec{p}+\frac{\vec{q}}{2}}}\right)\left(\Delta^{\vec{q}}_{\vec{k}+\vec{p}}\right)^{\tau\alpha,\tau'\beta}
\end{split}
\end{equation}
In the above, $\xi_{\vec{k}}^{\eta\sigma}=\epsilon_{\vec{k},\eta,\sigma}+\eta\mu_v+\sigma\mu_s$. We solve this linear equation for our various choices of normal states on a grid with 1657 $\vec{k}$ points and $\vec{q}$ points for each normal state.
\section{Different Normal States}\label{app:OtherNormalState}
\begin{figure}
    \centering
    \includegraphics[width=0.5\linewidth]{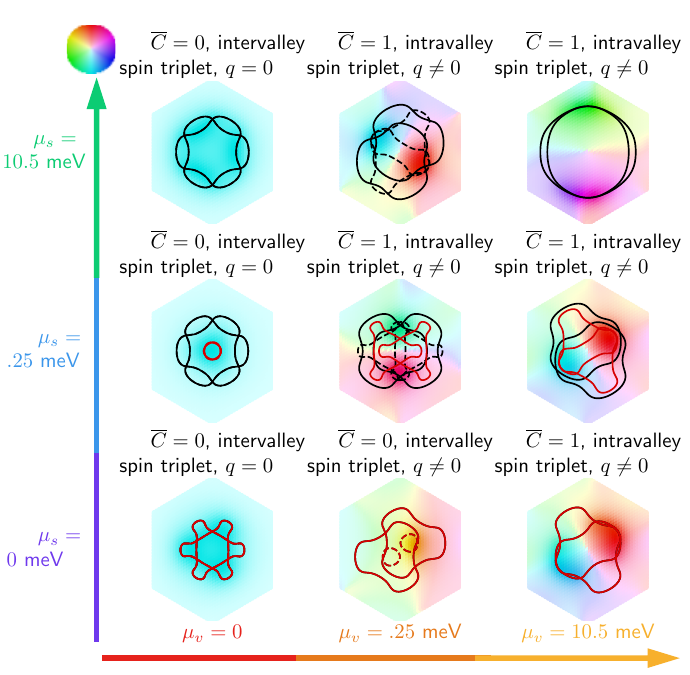}
    \caption{We plot the leading solution $\Delta_{\vec{k}}^{\vec{q}}$ for ferromagnetic fluctuations $(\rho^i=(s_x,s_y,s_z)\eta_0)$ as a function of varying amounts of spin and valley polarization in the normal state of superconductivity. We also plot the normal state Fermi surfaces for the $+$ (solid lines) and $-$ (dashed lines) valleys and $\uparrow$ (black) and $\downarrow$ (red) Fermi surfaces. We also use ``spin singlet" to refer to anti-symmetric pairing between spins and ``spin triplet" to refer a symmetric pairing state with respect to spin when $\mu_s\neq 0$ and SU(2)$_s$ spin symmetry is broken. We solve the linearied gap equation with displacement field $D=110$ meV. We observe a tendency towards a state with $\overline{C}=+1$ relative to the normal state with increasing valley polarization.}
    \label{fig:FM}
\end{figure}

In this appendix, we will study the impact of changing the underlying normal state on the leading superconducting instabilities. As shown in Fig.~\ref{fig:maxEig} and discussed in the main text, finite $\mu_v$ will generically result in finite $\vec{q}$ pairing. We next consider the impact of spin polarization on the leading instabilties by varying the amount of spin and valley polarization by adjusting $\mu_{s,v}$ in Eq.~\ref{Eq:kineticterm}. The results for ferromagentic fluctuations are shown in Fig.~\ref{fig:FM}. We find adding partial spin polarization ($\mu_s\neq 0$ such that both spin flavors both have active Fermi surfaces) pushes the leading instability to the $E$ state for smaller valley polarization as modeled by $\mu_v$.

\begin{figure}
    \centering
    \includegraphics[width=0.5\linewidth]{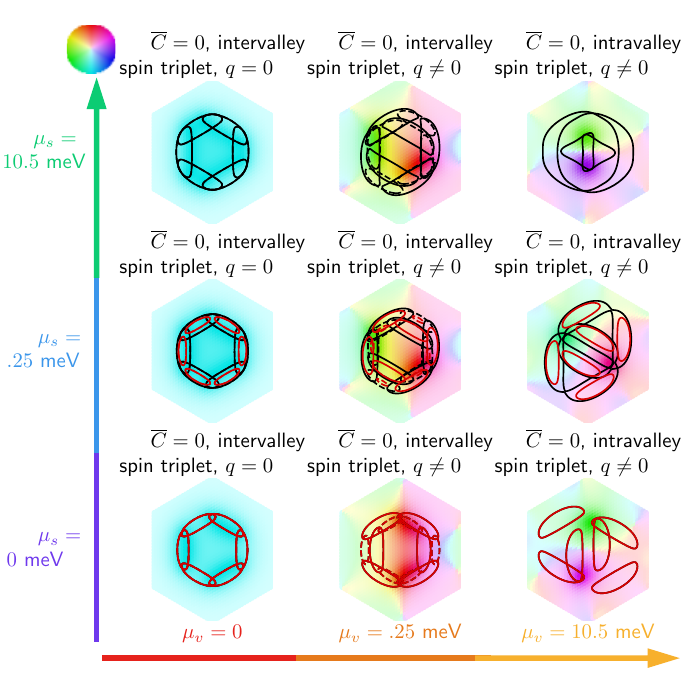}
    \caption{We plot the leading solution $\Delta_{\vec{k}}^{\vec{q}}$ for ferromagnetic fluctuations $(\rho^i=(s_x,s_y,s_z)\eta_0)$ as a function of varying amounts of spin and valley polarization in the normal state of superconductivity, for a normal state at the same doping of $n_e=0.6\times10^{12}$ cm$^-2$ but now at a higher displacement field $D=140$ meV. We plot the normal state Fermi surfaces for the $+$ (solid lines) and $-$ (dashed lines) valleys and $\uparrow$ (black) and $\downarrow$ (red) Fermi surfaces.}
    \label{fig:FM_normal2}
\end{figure}

\begin{figure}
    \centering
    \includegraphics[width=0.5\linewidth]{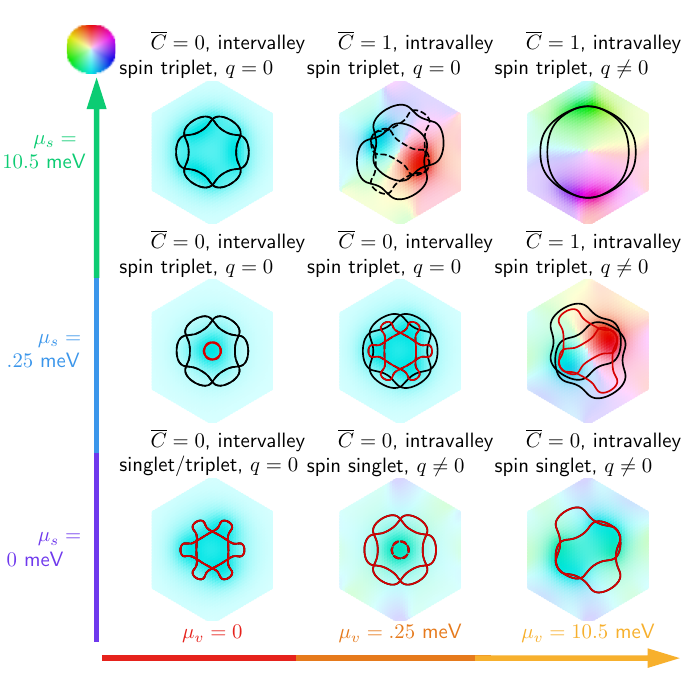}
    \caption{We plot the leading solution $\Delta_{\vec{k}}^{\vec{q}}$ for phonons $(\rho^i=s_0\eta_0)$ as a function of varying amounts of spin and valley polarization in the normal state of superconductivity. }
    \label{fig:PH}
\end{figure}

We then study ferromagnetic fluctuation mediated superconductivity for a normal state with a higher displacement field, such that the fully spin and valley polarized normal state has an annular Fermi surface. The results are shown in Fig.~\ref{fig:FM_normal2}. We find that the state always has $\overline{C}=0$ in agreement with the results of \cite{geier2024chiraltopologicalsuperconductivityisospin,yang2024topologicalincommensuratefuldeferrelllarkinovchinnikovsuperconductor,2025arXiv250217555Y,2025arXiv250305697M}.
Finally, we study the linearized gap equation for phonons $(\rho^i=s_0\eta_0)$ for varying amounts of spin and valley polarization. The results are shown in Fig.~\ref{fig:PH}.

\section{Fluctuations of Valley Polarized Order}\label{App:VP}
\begin{figure}
    \centering
    \includegraphics[width=0.5\linewidth]{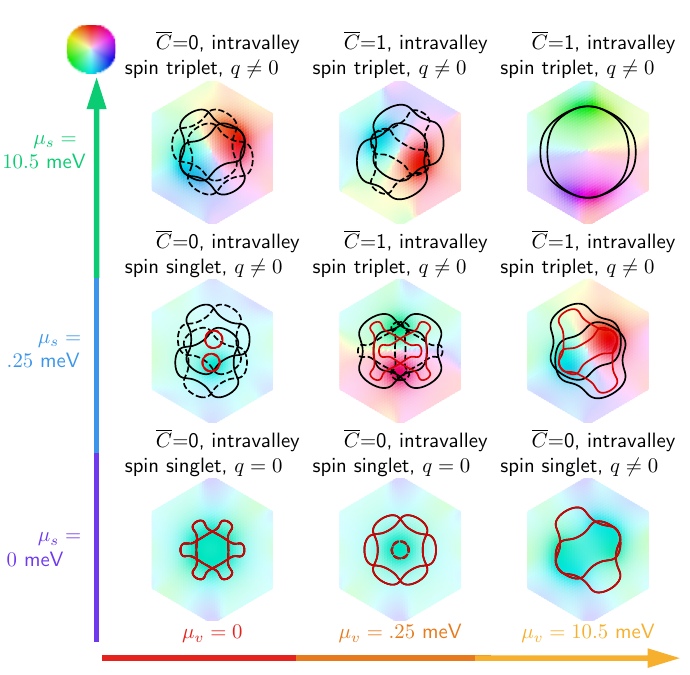}
    \caption{We plot the leading solution $\Delta_{\vec{k}}^{\vec{q}}$ for fluctuations of a valley polarized order $(\rho^i=s_0\eta_z)$ as a function of varying amounts of spin and valley polarization in the normal state of superconductivity. }
    \label{fig:VP}
\end{figure}

\begin{figure}
    \centering
    \includegraphics[width=.5\linewidth]{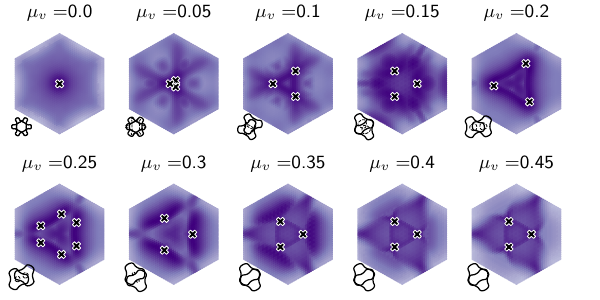}
    \caption{We plot the leading eigenvalue $\lambda(\vec{q})$ corresponding to each $\Delta_{\vec{k}}^{\vec{q}}$ as a function of center of mass momentum $\vec{q}$, for varying amounts of valley polarization and $\mu_s=0$. Maxima are marked with a x marker.}
    \label{fig:maxEig}
\end{figure}

In this appendix we discuss the leading superconducting instabilities for the case of valley polarized fluctuations (ie $\rho=\eta_zs_0$). Our results are shown in Fig.~\ref{fig:VP}. We find an intravelly state is favored for all values of $\mu_v$ and $\mu_s$, with a transition to a state with $\overline{C}=1$ occurring for sufficient spin polarization. A finite-$\vec{q}$ state is also favored in the spin polarized case even when $\mu_v=0$, as a consequence of disfavored pairing between Kramers partners for this type of interaction. 

\section{Microscopic evaluation of Landau coefficients}
\label{app: Landau theory}
In this appendix, we derive and discuss the microscopic expressions for the coefficients of the Landau theory in Eq.~\eqref{eq: Landau th}. We start from a continuum model of spinless electrons with a trigonal warping term (parametrized by $w$) coupled to a $p+ip$-wave superconducting order parameter:
\begin{equation}
    \mathcal{H}= \sum_{\vec{k}} \xi_{\vec{k}}\, c^\dagger_{\vec{k}} c_{\vec{k}} + \sum_{\vec{k},\vec{q}}\left[e^{i\theta_{\vec{k}}}\, c_{{\vec{q}/2-\vec{k}}} c_{{\vec{q}/2+\vec{k}}}\, \Delta^*_{\vec{q}}+\text{h.c.}\right]\,,
\end{equation}
with $\xi_{\vec{k}}=\frac{\vec{k}^2}{2m}\left[1+w\cos(3\theta_{\vec{k}})\right]-\mu$, $\mu$ being the chemical potential and $m$ the electron mass, and $e^{i\theta_{\vec{k}}}=\frac{k_x+ik_y}{|\vec{k}|}$. We write $\Delta_{\vec{q}}=\sum_{n=1}^3 \phi_n\,\delta_{\vec{q},\vec{q}_n}$, where $\vec{q}_n$ are three $C_{3z}$-related wave vectors given by $\vec{q}_1=|q|(1,0)$, $\vec{q}_2=|q|(-1/2,\sqrt{3}/2)$, and $\vec{q}_3=|q|(-1/2,-\sqrt{3}/2)$. Integrating out the electrons and expanding the effective action up to fourth order in $\phi_{n}$, we obtain Eq.~\eqref{eq: Landau th}. The coefficients $u$ and $v$ are given by
\begin{subequations}
    \begin{align}
        &u = I_1\,\\
        & v = 2I_2 - 2I_1\,,
    \end{align}
\end{subequations}
where the integrals $I_1$ and $I_2$ read
\begin{subequations}
    \begin{align}
        I_1=\int_{\vec{k}} \Big[ &2\,\frac{n_F(-\xi_{\vec{q}_1-\vec{k}})-n_F(\xi_{\vec{k}})}{(\xi_{\vec{k}}+\xi_{\vec{q}_1-\vec{k}})^3}+\frac{n_F'(-\xi_{\vec{q}_1-\vec{k}})+n_F'(\xi_{\vec{k}})]}{(\xi_{\vec{k}}+\xi_{\vec{q}_1-\vec{k}})^2}\Big]\,,\\
        I_2=\int_{\vec{k}} \Big[&
        -\frac{(2\eo+\etw+\eth)}{(\eo+\etw)^2(\eo+\eth)^2}n_F(\eo)
        +\frac{n_F(-\eth)}{(\eo+\eth)^2(\etw-\eth)}\nonumber\\
        &-\frac{n_F(-\etw)}{(\eo+\etw)^2(\etw-\eth)}
        +\frac{n_F'(\eo)}{(\eo+\etw)(\eo+\eth)}
        \Big]\,,
    \end{align}
\end{subequations}
with $n_F(x)=1/(e^{x/T}+1)$ the Fermi distribution function at temperature $T$ and $n_F'(x)=\frac{d n_F}{dx}$. In the above equations the momentum integrals have to be intended as $\int_{\vec{k}}=\int_0^\Lambda \frac{|\vec{k}|d|\vec{k}|}{2\pi}\int_{0}^{2\pi}\frac{d\theta_{\vec{k}}}{2\pi}$. In the numerical evaluation of the coefficients in Fig.~\ref{fig:fig3} we have chosen $\Lambda=10k_F$ with $k_F=\sqrt{2m\mu}$ and a temperature of $T=E_F/4$, with $E_F=\mu$.

\section{Chern Number}\label{app:Chern}
In this appendix, we will describe how the Chern number of each superconducting state is computed within the continuum model description of App.~\ref{App:LGE}. For each pairing state, we construct the mean-field Hamiltonian:
\begin{equation}\label{Eq:HBdG}
    H_{BdG}=\sum_{\vec{k}}\Psi_{\vec{k}}^\dagger 
    \begin{pmatrix}\tilde{\epsilon}_{\vec{k},\vec{q}}&\Delta_{\vec{k}}^{\vec{q}}\\\left(\Delta_{\vec{k}}^{\vec{q}}\right)^\dagger & -\tilde{\epsilon}_{-\vec{k},\vec{q}}
    \end{pmatrix}
    \Psi_{\vec{k}} 
\end{equation}
Where $\Psi_{\vec{k}}$ is a vector defined for each $\vec{k}$ on half the Brillouin zone as:
\begin{equation}
    \Psi_{\vec{k}}=\left(c_{\vec{k}+\frac{\vec{q}}{2},\uparrow,+},c_{\vec{k}+\frac{\vec{q}}{2},\downarrow,+},c_{\vec{k}+\frac{\vec{q}}{2},\uparrow,-},c_{\vec{k}+\frac{\vec{q}}{2},\downarrow,-},c^\dagger_{-\vec{k}+\frac{\vec{q}}{2},\uparrow,+},c^\dagger_{-\vec{k}+\frac{\vec{q}}{2},\downarrow,+},c^\dagger_{-\vec{k}+\frac{\vec{q}}{2},\uparrow,-},c^\dagger_{-\vec{k}+\frac{\vec{q}}{2},\downarrow,-}\right)^T
\end{equation}
and $\tilde{\epsilon}_{\vec{k},\vec{q}}$ is the diagonal matrix constructed from the dispersions of each spin and valley flavor:
\begin{equation}
    \tilde{\epsilon}_{\vec{k},\vec{q}}=\text{Diag}\left(
        \xi_{\vec{k}+\frac{\vec{q}}{2},\uparrow,+},\xi_{\vec{k}+\frac{\vec{q}}{2},\downarrow,+},\xi_{\vec{k}+\frac{\vec{q}}{2},\uparrow,-},\xi_{\vec{k}+\frac{\vec{q}}{2},\downarrow,-}\right)
\end{equation} 
and $\Delta_{\vec{k}}^{\vec{q}}$ is the matrix defined in Eq.~\ref{Eq:deltaDef}. For a given band of Eq.~\ref{Eq:HBdG} with corresponding set of eigenvectors at each $\vec{k}$ point labeled as $a_{\vec{k}}$, we compute the Chern number of the superconducting state as:
\begin{equation}
    C=\frac{1}{2\pi}\int_{BZ} d^2\vec{k}\quad \Omega^{SC}(\vec{k}) 
\end{equation}
where $\Omega^{SC}(\vec{k})$ is the Berry curvature of the superconducting state, and is defined as:
\begin{equation}
\begin{split}
\Omega^{SC}_{\vec{k}}&=\text{Im}\left[\text{log}\left(U_{\vec{k},\vec{k}+\delta k\vec{\hat{x}}}U_{\vec{k}+\delta k\vec{\hat{x}},\vec{k}+\delta k\vec{\hat{x}}+\delta k \vec{\hat{y}}}U_{\vec{k}+\delta k\vec{\hat{x}}+\delta k \vec{\hat{y}},\vec{k}+\delta k \vec{\hat{y}}}U_{\vec{k}+\delta k \vec{\hat{y}},\vec{k}}\right)\right] \\&\text{where} \quad U_{\vec{k},\vec{k'}}=a^\dagger_{\vec{k}}\begin{pmatrix}F(\vec{k}+\frac{\vec{q}}{2},\vec{k'}-\vec{k}) & 0 \\ 0 & F^*(-\vec{k}+\frac{\vec{q}}{2},-\vec{k'}+\vec{k})
    \end{pmatrix}a_{\vec{k'}}
\end{split}
\end{equation}

\begin{equation}
    U_{\vec{k},\vec{k'}}=a^\dagger_{\vec{k}}\begin{pmatrix}F(\vec{k}+\frac{\vec{q}}{2},\vec{k'}-\vec{k}) & 0 \\ 0 & F^*(-\vec{k}+\frac{\vec{q}}{2},-\vec{k'}+\vec{k})
        
    \end{pmatrix}a_{\vec{k'}}\qquad \Omega(\vec{k})=\text{Im}\left[\text{log}\left(U_{\vec{k}_1,\vec{k}_2}U_{\vec{k}_2,\vec{k}_3}U_{\vec{k}_3,\vec{k}_4}U_{\vec{k}_4,\vec{k}_1}\right)\right]
\end{equation}
Our continuum model has a cutoff $|\vec{\Lambda}|>k_F$ within each valley (in our computations we use a hexagonal momentum space grid with a $\vec{k}$ dependent cutoff, though this does not affect the following argument) and we may break the berry curvature into two contributions, one for integrand with $|\vec{k}|<|\vec{\Lambda}|$ and one for integrand with $|\vec{k}|>|\vec{\Lambda}|$, such that we may write:
\begin{equation}
    C_{SC}=\frac{1}{2\pi}\int_{|\vec{k}|<|\vec{\Lambda}|} d^2{\vec{k}}\quad\Omega^{SC}_{\vec{k}}+\frac{1}{2\pi}\int_{|\vec{k}|>|\vec{\Lambda}|} d^2{\vec{k}}\quad\Omega^{SC}_{\vec{k}}
\end{equation}
We note that for cases where the normal state is totally valley polarized, we are taking the cutoff $|\vec{\Lambda}|$ to be within only the single, active valley. For cases where the pairing is valley polarized but the normal state still has active Fermi surfaces in both valleys, we compute the Chern number only for the valley in which the pairing is active (although our definition of $C$ is still well-defined and quantized in this case, we note that quantized response would not be observed due to the presence of the active Fermi surface in the other valley). For cases with an intervalley pairing, we take a cutoff $|\vec{\Lambda}|$ in both valleys and compute the Chern number.

If $|\Delta_{\vec{k}}^{\vec{q}}|$ is sufficiently small for $|\vec{k}|>|\vec{\Lambda}|$, the second contribution may be rewritten as: 
\begin{equation}
    \frac{1}{2\pi}\int_{|\vec{k}|>|\vec{\Lambda}|} d^2{\vec{k}}\quad\Omega^{SC}_{\vec{k}}=-\frac{1}{2\pi}\int_{|\vec{k}|>|\vec{\Lambda}|} d^2{\vec{k}}\quad\Omega^{\phi}_{\vec{k}}=\frac{1}{2\pi}\int_{|\vec{k}|<|\vec{\Lambda}|} d^2{\vec{k}}\quad\Omega^{\phi}_{\vec{k}}
\end{equation}
where $\Omega^{\phi}_{\vec{k}}$ is the Berry curvature of the normal state Bloch wavefunction. In the above we have used that the Berry curvature integrated over the entire band (including both valleys of the continuum model) must be 0. In this case, we have an expression for the Berry curvature defined for integrands only within our momentum cutoff:
\begin{equation}
    C_{SC}=\frac{1}{2\pi}\int_{|\vec{k}|<|\vec{\Lambda}|} d^2{\vec{k}}\quad\Omega^{SC}_{\vec{k}}+ \frac{1}{2\pi}\int_{|\vec{k}|<|\vec{\Lambda}|} d^2{\vec{k}}\quad\Omega^{\phi}_{\vec{k}}
\end{equation}

\section{3-q state gap}
\begin{figure}
    \centering
    \includegraphics[width=\linewidth]{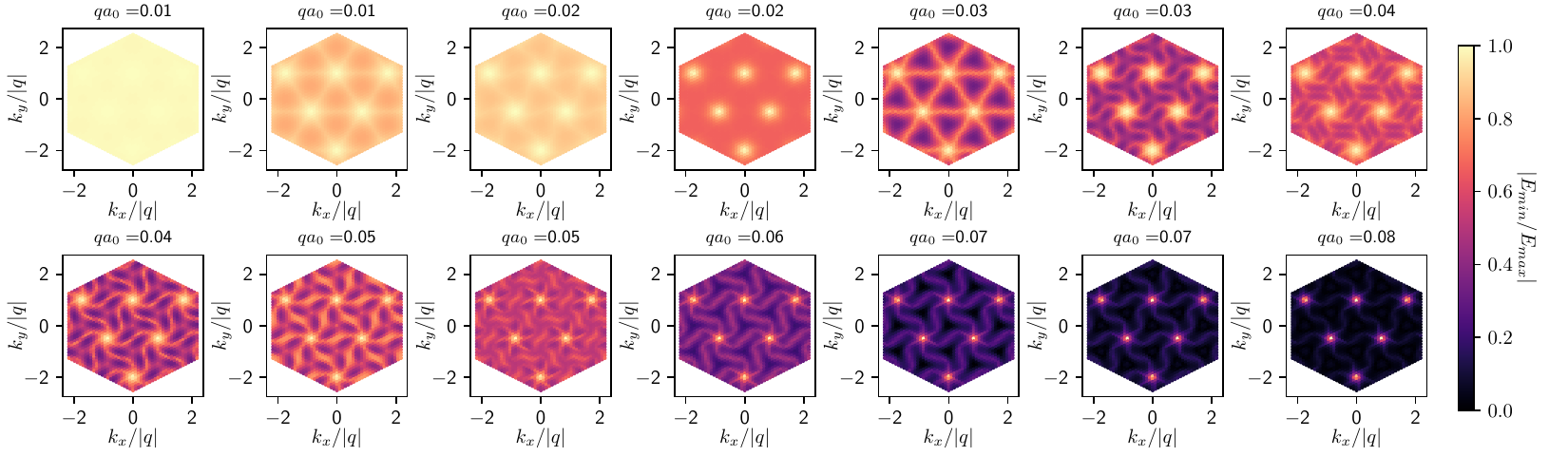}
    \caption{We plot the energy of the lowest band of the mean-field Hamiltonian studied for the 3-$\vec{q}$ state at $|\Delta|=0.1$ meV as  function of $|\vec{q}|$, where $\vec{q}$ lies along a high symmetry contour in momentum space.}
    \label{fig:moiregap}
\end{figure}

In this appendix, we discuss how the minimum excitation gap is computed for the 3-$\vec{q}$ state for a finite $\vec{q}$, momentum independent pairing $\Delta_{\vec{k}}^{\vec{p}}=|\Delta|$. For simplicity, we assume spin singlet pairing ($\mu_s=0$) and only a single active valley with dispersion $\xi_{\vec{k}}$ (we will suppress valley indices in our discussion going forward). We then consider adding a term with the following form in real space:
\begin{equation}
    \mathcal{H}_{pair}=\sum_{\vec{i}}\sum_jc^\dagger_{\vec{i},\uparrow}c^\dagger_{\vec{i},\downarrow}e^{i\vec{Q}_j\cdot\vec{r}}+\text{h.c.}
\end{equation}
where $j$ is summed over the three $C_3$ related $\vec{Q}_i$ vectors. Fourier transforming to momentum space, we can compute the spectrum by defining the basis:
\begin{equation}
    \Psi_{\vec{k}}=\left(c_{\vec{k},\uparrow},c^\dagger_{-\vec{k}+\vec{Q}_1,\downarrow},c^\dagger_{-\vec{k}+\vec{Q}_2,\downarrow},c^\dagger_{-\vec{k}+\vec{Q}_3,\downarrow},...\right)
\end{equation}
where the ellipses denotes the continued definition of the basis, alternating between sets of spin up creation operators and sets of spin down annihilation operators such that each subsequent set of either creation or annihilation operators have an element at a momenta equal to the negative momentum plus $\vec{Q}_j$ of an element in the previous set. We set a cutoff by fixing the number of these ``shells", with each shell defined by a set of momenta at fixed distance from the origin when $\vec{k}=0$). Within this basis, the Hamiltonian we study is constructed as
\begin{equation}\label{finiteQModel}
    \mathcal{H}=\sum_{\vec{k}}\Psi^\dagger_{\vec{k}}\begin{pmatrix}
        \xi_{\vec{k}}& \Delta& \Delta & \Delta &\dots \\ \Delta & -\xi_{-\vec{k}+\vec{Q_1}} & 0 & 0& \\ \Delta& 0  & -\xi_{-\vec{k}+\vec{Q_2}} & 0 & \\ \Delta& 0  & 0 & -\xi_{-\vec{k}+\vec{Q_3}}&  \\ \vdots & & & &\ddots
    \end{pmatrix}\Psi_{\vec{k}}
\end{equation}
where again the ellipses denote continued dispersion on the diagonal with alternating sign for dispersions at momenta in each shell and $\Delta$ continuing on the off diagonal coupling dispersions at momenta which sum to $\vec{Q}_j$. The absolute value of the energy of the state closest to zero energy obtained from diagonalizing Eq.~\ref{finiteQModel} is shown in Fig.~\ref{fig:moiregap} for 17 shells of basis vectors as a function of $q=|\vec{Q}_i|$. The number of shells necessary to achieve convergence is on the order of $k_F/q$. We note the non-monotonic behavior as a function of $q$ shown in Fig.~\ref{fig:fig3} originates from the coupling between different Fermi surfaces at different $q$ as induced by $\Delta$.
\section{Sign of "valley-independent Chern number" $\overline{C}$}\label{app:chirality}
In this appendix, we will discuss on more general grounds how the underlying Berry curvature of the normal-state Bloch states determines the sign of the Chern number $C$ of the superconductor (and, thus, $\overline{C}$) for the interactions we study. Our starting point is the linearized gap equation. In order to consider the simplest tractable case, we will set the COM momentum $\vec{q}\rightarrow 0$. 
We also assume a finite (but not necessarily large) degree of both spin and valley polarization and neglect the cases of pairing between opposite valleys and spins, assuming that this scenario is energetically disfavored in a (potentially partially) polarized normal state. Under these assumptions, the linearized gap equation takes the form
\begin{equation}
    \Delta_{\vec{k}}^{\gamma}\sim \sum_{\vec{k}'}\mathcal{V}^{\gamma,\delta}_{\vec{k},\vec{k}'}(T)F^{\gamma,\delta}_{\vec{k},\vec{k}'}F_{-\vec{k},-\vec{k}'}^{\gamma\delta}\Delta^{\delta}_{\vec{k}'}, \label{FirstFormOfGapEquation}
\end{equation}
where $\mathcal{V}^{\gamma,\delta}_{\vec{k},\vec{k}'}(T)$ comprises both the interaction potential as well as thermal and kinetic factors.
In the above $\gamma,\delta$, are indices in the four-dimensional space of spin and valley. In general, the form factors $F$ in terms of the normal-state Bloch wavefunctions $\ket{u_\gamma(\vec{k})}$ are $F_{\vec{k},\vec{k}'}^{\gamma\gamma'}=\bra{u_\gamma(\vec{k})}\rho^{\gamma\gamma'} \ket{u_{\gamma'}(\vec{k'})}$, where, for given $\gamma,\gamma'$, $\rho^{\gamma\gamma'}$ is a potentially non-trivial matrix in the remaining flavor quantum numbers (e.g., sublattices and layers). For concreteness, we will here first focus on the special case $\rho^{\gamma\gamma'} = \rho_\gamma \delta_{\gamma,\gamma'} \mathbbm{1}$, $\rho_\gamma \in \mathbb{R}$,  relevant to the scenarios discussed in the main text. The form factors then simply become $F_{\vec{k},\vec{k'}}^{\gamma\gamma'} = \delta_{\gamma\gamma'} \rho_\gamma \braket{u_\gamma(\vec{k})|u_{\gamma'}(\vec{k}')}$.

We next further assume that $\Delta_{\vec{k}}$ will only be non-zero in the immediate vicinity of the Fermi surface. Within this thin strip around the Fermi surface, it is then natural to approximate $\Delta_{\vec{k}}$ and the Bloch states/interaction matrix elements to only depend on the position on the Fermi surface, which we parametrize with an angular-like variable $\varphi$. 
Then \equref{FirstFormOfGapEquation} can be compactly written as
\begin{equation}
    \Delta^{\gamma}_{\varphi}\sim \int d\alpha \,V^\gamma_{\alpha}(T) F^{\gamma}_{\varphi,\varphi+\alpha}F^{\gamma}_{\varphi+\pi,\varphi+\alpha+\pi}\Delta^{\gamma}_{\varphi+\alpha}. \label{AngularFormOfGapEquation}
\end{equation}
Since different $\gamma$ do not couple, in order to solve the linearized gap equation, it is sufficient to focus on the values of $\gamma$ with the largest transition temperature. This is why we suppressed the superscript $\gamma$ in our description of this analysis in the main text.  

To gain intuitive insights, let us further assume that the effective interaction is sufficiently short-ranged in momentum space (long-ranged in real space) such that the integrand in \equref{AngularFormOfGapEquation} can be expanded in powers of $\alpha$. 
We start by expanding the form factors to leading order in $\alpha$:
\begin{equation}
    F^{\gamma}_{\varphi,\varphi+\alpha}=  \rho_\gamma \left( 1+\alpha \braket{u_\gamma(\varphi)|\partial_\varphi u_\gamma(\varphi)} \right)+\mathcal{O}(\alpha^2)
\end{equation}
In the above expression, we can identify the second term in this expansion as proportional to the normal-state Berry connection along the Fermi surface since 
\begin{equation}
     i\braket{u_\gamma(\varphi)| \partial_\varphi u_\gamma(\varphi)} = k_F(\varphi) \hat{\vec{e}}_\varphi \cdot i \braket{ u_\gamma(\vec{k}) | \vec{\nabla}_{\vec{k}} u_\gamma(\vec{k}) } =: k_F(\varphi) \mathcal{A}_{\varphi}^{\gamma},
\end{equation}
where $k_F(\varphi)$ is the Fermi wavevector at point $\varphi$ on the Fermi surface and $\hat{\vec{e}}_\varphi$ the unit vector tangential to the Fermi surface. 
To leading order in $\alpha$ the linearized gap equation reads
\begin{equation}
\begin{split}
    \Delta^{\gamma}_{\varphi}&= \rho_\gamma^2 \int d\alpha \,V_{\alpha}(T) \left[1-i\alpha k_F(\varphi) \left(\mathcal{A}_{\varphi}^{\gamma}+ \mathcal{A}_{\varphi+\pi}^{\gamma}\right)\right]\Delta_{\varphi+\alpha}^{\gamma} \label{LinGapinAlphaForm}
\end{split}
\end{equation}
We then similarly expand $\Delta^{\gamma}_{\varphi+\alpha}=\Delta^{\gamma}_{\varphi}+\alpha\partial_\varphi\Delta^\gamma_{\varphi}+\frac{1}{2}\alpha^2\partial_\varphi^2\Delta^\gamma_{\varphi} +\mathcal{O}(\alpha^3)$. Keeping all terms up to order $\alpha^2$ in the integrand, defining $\chi^\gamma_n := \rho_\gamma^2\int d \alpha \, V_{\alpha}(T) \alpha^n$, and making the common assumption that the effective interaction is even such that $\chi_1=0$, \equref{LinGapinAlphaForm} becomes
\begin{equation}
\begin{split}\label{Eq:ApproxGap}
    \Delta^{\gamma}_{\varphi}&\sim  \rho_\gamma^2 \int d\alpha \,V_{\alpha}(T)\left[\Delta^{\gamma}_\varphi+\frac{1}{2}\alpha^2\partial_\varphi^2\Delta^{\gamma}_\varphi\right]-i \rho_\gamma^2 k_F(\varphi) \int d\alpha \left[\alpha (\mathcal{A}^\gamma_{\varphi}+ \mathcal{A}^\gamma_{\varphi+\pi})\right]\alpha\partial_\varphi\Delta^\gamma_{\varphi}\\
    &= \chi^\gamma_0 \Delta^{\gamma}_\varphi-i k_F(\varphi) \chi^\gamma_2  \left(\mathcal{A}^\gamma_\varphi +\mathcal{A}^\gamma_{\varphi+\pi}\right) \partial_\varphi \Delta^{\gamma}_\varphi + \frac{1}{2}\chi^\gamma_2 \partial^2_\varphi\Delta^{\gamma}_{\varphi}.
\end{split}
\end{equation}
So the linearized gap equation can be stated as $\Delta^\gamma_\varphi = \hat{K}^\gamma \Delta^\gamma_\varphi$ with differential operator 
\begin{equation}
    \hat{K}^\gamma = \chi_0 -i \chi^\gamma_2 k_F(\varphi) \left(\mathcal{A}^\gamma_\varphi +\mathcal{A}^\gamma_{\varphi+\pi}\right) \partial_\varphi + \frac{1}{2}\chi_2 \partial^2_\varphi,  \label{KOperator}
\end{equation}
as stated in the main text. First, consider the limit without the term proportional to the Berry connection in $\hat{K}^\gamma$. If $\Delta^\gamma_\varphi$ is a solution of the gap equation, $\Delta^\gamma_{-\varphi}$, which has the opposite chirality, will always be degenerate with it (i.e., we have a non-chiral state or pairing in a multi-dimensional irreducible representation). The second term in \equref{KOperator} splits the degeneracy between these solutions and, thereby, selects a chirality. To see this, let us assume that $\Delta^\gamma_\varphi$ solves the full gap equation. Then the difference $\Delta \lambda$ in eigenvalue between this solution and  $\Delta^\gamma_{-\varphi}$ reads as
\begin{equation}
    \Delta \lambda = 2 \chi^\gamma_2 \int d\varphi \, k_F(\varphi)  \left(\mathcal{A}^\gamma_\varphi +\mathcal{A}^\gamma_{\varphi+\pi}\right) \left[-i \left(\Delta^{\gamma}_\varphi\right)^* \partial_\varphi \Delta^{\gamma}_\varphi\right] = 2 |\Delta_0|^2 \chi^\gamma_2 \int d\varphi \, k_F(\varphi)  \left(\mathcal{A}^\gamma_\varphi +\mathcal{A}^\gamma_{\varphi+\pi}\right) \partial_\varphi \alpha 
\end{equation}
where we wrote $\Delta^{\gamma}_\varphi = |\Delta_0| e^{i\alpha(\varphi)}$ in the second equality. Taking an attractive interaction ($\chi_2>0$), this shows that the leading superconducting solution will be the one that maximizes a ``weighted'' version, $w_{\mathcal{A}} := \int d\varphi \, k_F(\varphi)  \left(\mathcal{A}^\gamma_\varphi +\mathcal{A}^\gamma_{\varphi+\pi}\right) \partial_\varphi \alpha$, of the winding number $w = \int d\varphi \,  \partial_\varphi \alpha \in \mathbbm{Z}$ of the superconducting order parameter. Assuming that (there is a gauge where) $\mathcal{A}^\gamma_\varphi +\mathcal{A}^\gamma_{\varphi+\pi}$ has a fixed sign (as is the case in rhombohedral graphene), we see that
\begin{equation}
    \sign( \partial_{\varphi} \alpha) =  \sign\left(\mathcal{A}^\gamma_\varphi +\mathcal{A}^\gamma_{\varphi+\pi}\right) = \text{const}., \label{RelationBetweenSigns}
\end{equation}
linking the normal-state chirality to the chirality of the resulting superconductor. 

Finally, it is left to connect the left and right hand side of \equref{RelationBetweenSigns} to the sign of $C$ and $\Phi_N$, respectively, which will determine the sign of the valley independent Chern number $\bar{C}$. To this end, we first choose a gauge where the Bloch states $\ket{u_\gamma(\vec{k})}$ are smooth in the momentum-space area $A$ enclosed by the Fermi surface. It then holds
\begin{equation}
    \Phi_N = \int_A d^2\vec{k} \, \Omega_{\vec{k}} = \int_A d^2\vec{k} \, \vec{\nabla}_{\vec{k}} \times i \braket{ u_\gamma(\vec{k}) | \vec{\nabla}_{\vec{k}} u_\gamma(\vec{k}) }  = \oint_{\partial A} d\vec{k}  \, i \braket{ u_\gamma(\vec{k}) | \vec{\nabla}_{\vec{k}} u_\gamma(\vec{k}) } = \int d\varphi \, k_F(\varphi) \mathcal{A}_{\varphi}^{\gamma},
\end{equation}
where $\Omega_{\vec{k}} = -2 \text{Im} \braket{\partial_{k_x} u_\gamma | \partial_{k_y } u_\gamma }$ is the normal-state Berry curvature. This immediately implies
\begin{equation}
    \sign(\Phi_N) = \sign\left(\mathcal{A}^\gamma_\varphi +\mathcal{A}^\gamma_{\varphi+\pi}\right). \label{SignPhiN}
\end{equation}
As shown in \cite{TheSTMPaper}, in this gauge, the Chern number of the superconductor can just be obtained by integrating the ``naive'' superconducting Berry curvature,
\begin{equation}
        \widetilde{\Omega}_{\text{SC}}(\vec{k}) = \frac{1}{2 |\vec{g}_{\vec{k}}|^3} \vec{g}_{\vec{k}} \cdot (\partial_x \vec{g}_{\vec{k}} \times \partial_y \vec{g}_{\vec{k}}), \quad \vec{g}_{\vec{k}} = (\Re \Delta_{\vec{k}}, -\Im \Delta_{\vec{k}}, \xi_{\vec{k}} )^T
\end{equation}
over the low-energy theory, where $\xi_{\vec{k}}$ is the normal-state dispersion. It is straightforward to see that within the low-energy theory
\begin{equation}
    \sign (C) = \sign \left( \widetilde{\Omega}_{\text{SC}}(\vec{k}) \right) = -\sign (\xi_{\vec{k}=0}) \sign( \partial_{\varphi} \alpha), \label{SignC}
\end{equation}
where $\sign (\xi_{\vec{k}=0}) < 0$ and $\sign (\xi_{\vec{k}=0}) > 0$ for an electron-like and hole-like pocket, respectively. Combining Eqs.~(\ref{RelationBetweenSigns}), (\ref{SignPhiN}), and (\ref{SignC}), we come to our final conclusion:
\begin{equation}
    \sign (C) = -\sign (\xi_{\vec{k}=0}) \sign(\Phi_N) \quad \text{and thus} \quad \sign (\bar{C}) = -\sign (\xi_{\vec{k}=0}).
\end{equation}
For the conduction band in rhombohedral graphene, we have $\sign (\xi_{\vec{k}=0}) <0$ and thus $\sign (\bar{C}) = 1$, in line with our numerics. We close by pointing out that one could just as well have defined $\Phi_N$ with an extra minus sign for a hole-like pocket, yielding $\sign(\bar{C}) = 1$ in both cases.

\section{Behavior for general polarization}\label{AppH}
\begin{figure}
    \centering
    \includegraphics[width=0.5\linewidth]{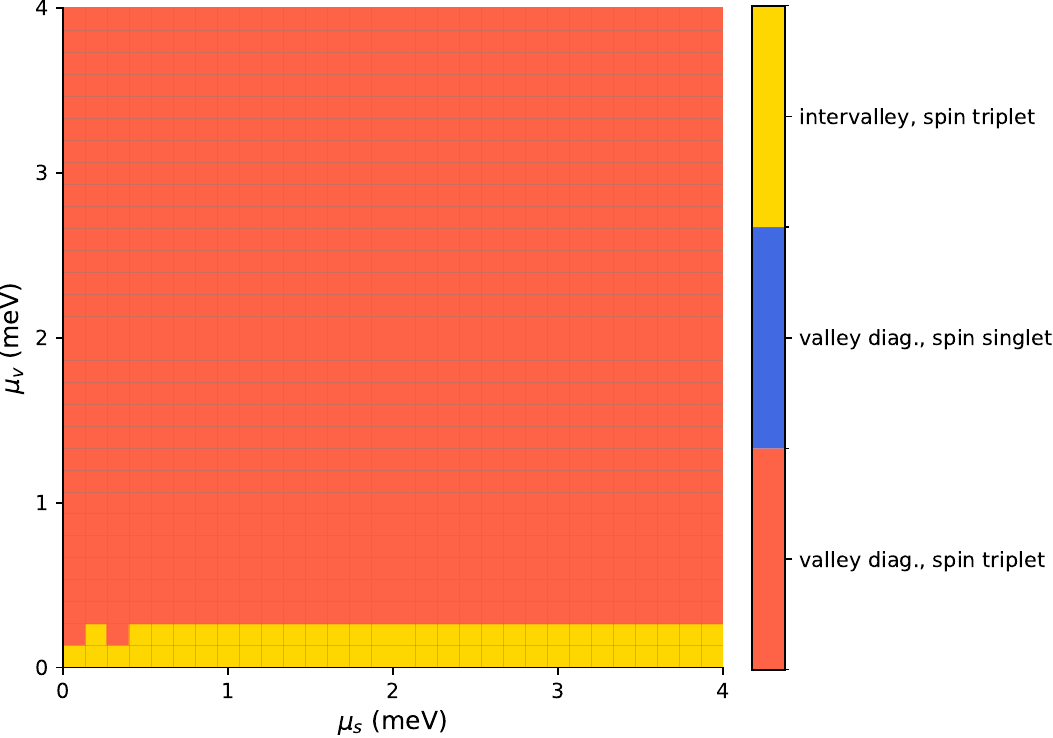}
    
    \caption{Phase diagram as computed from the leading eigenvalue of the $\vec{q}=0$ linearized gap equation for ferromagnetic fluctuation-mediated interactions as a function of $\mu_s$ and $\mu_v$.}
    \label{fig:Ferrophase}
\end{figure}

\begin{figure}
    \centering
    \includegraphics[width=0.5\linewidth]{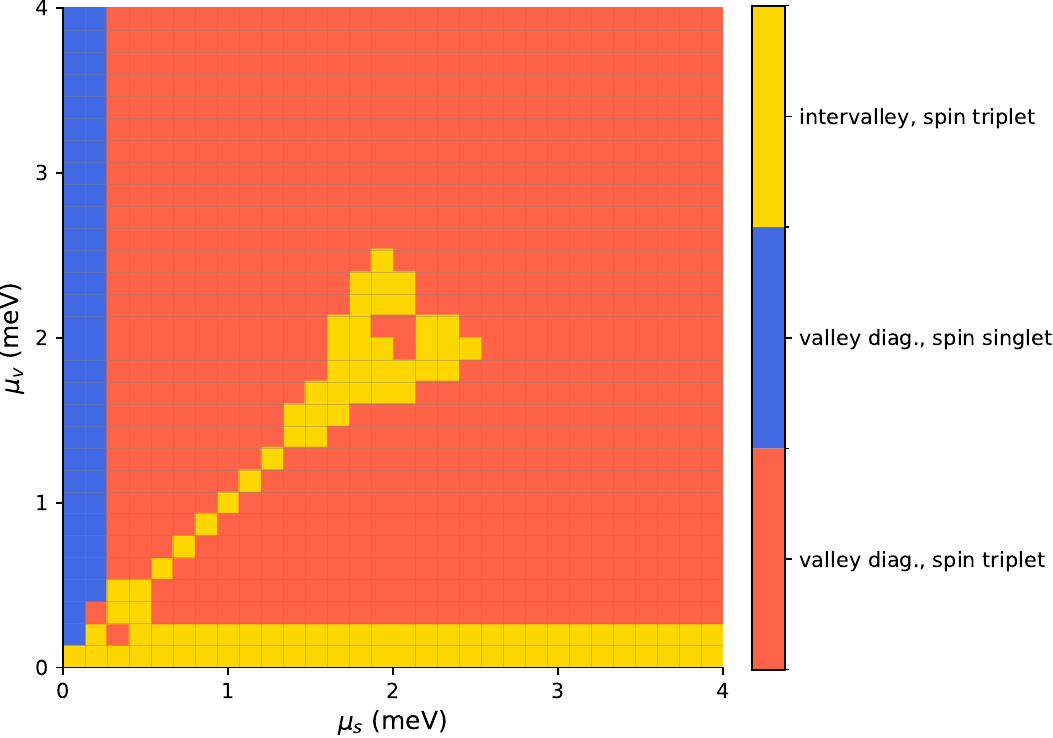}
    
    \caption{Phase diagram as computed from the leading eigenvalue of the $\vec{q}=0$ linearized gap equation for phonon-mediated interactions as a function of $\mu_s$ and $\mu_v$.}
    \label{fig:phophase}
\end{figure}

In this this appendix, we compute phase diagrams as a function of fine-grained values of the chemical potentials $\mu_s$ and $\mu_v$, allowing us to demonstrate general features of the leading superconducting pairing beyond the representative values studied in the remainder of our paper. In particular, we compute the leading eigenvalue of the gap equation for phonon-mediated and ferromagnetic fluctuation mediated superconductivity. In order to simplify the calculation, we take a smaller system size (an 18$\times$18 momentum space grid) and restrict our solutions to $\vec{q}=0$ only. The results are shown in Fig.~\ref{fig:Ferrophase} for ferromagnetic fluctuations and in Fig.~\ref{fig:phophase} for phonons.

In both cases, we observe a transition from a valley off-diagonal, topologically trivial phase to an intra-valley, topologically non-trivial, spin triplet phase as $\mu_v$ increases, consistent with the observation that fully polarizing the valley degree of freedom induces a topologically non-trivial phase. For the case of phonon-mediated superconductivity, we also observe a (topologically trivial) spin-singlet phase persist for small values of $\mu_s$. 

\begin{figure}[htbp]
    \centering
    \begin{subfigure}{0.45\textwidth}
        \centering
        \includegraphics[width=\linewidth]{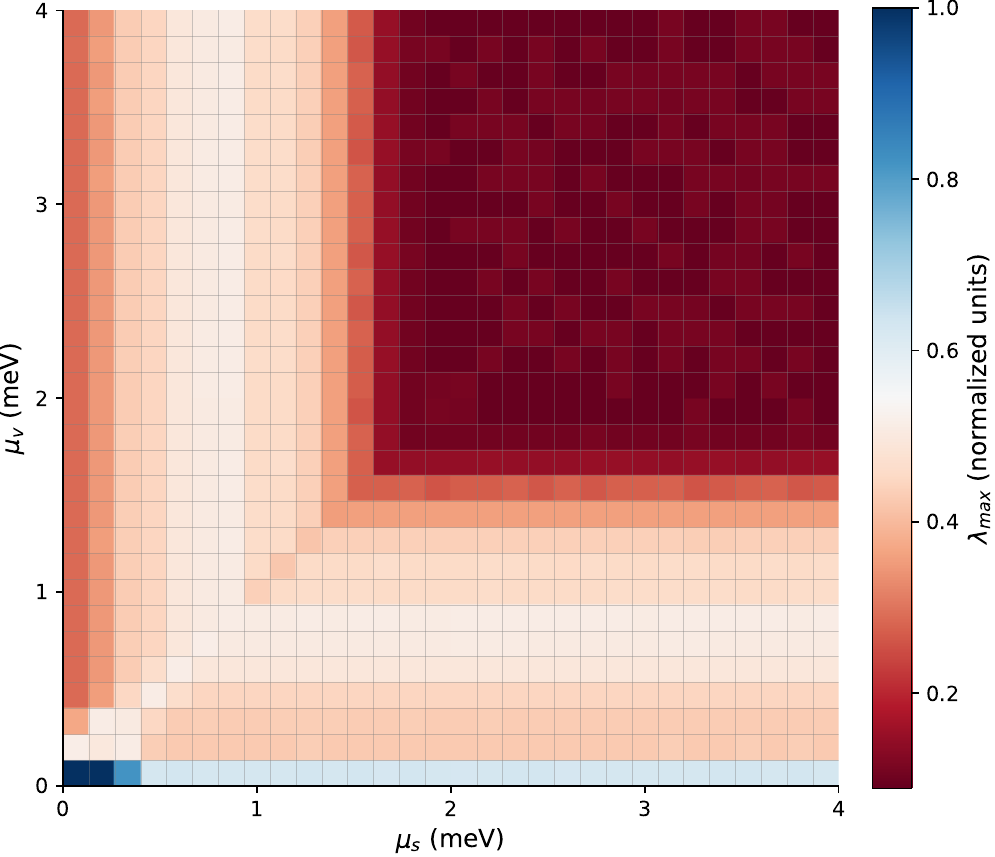}
        \caption{We show the leading eigenvalue $\lambda_{max}$ of the linearized gap equation for phonon-mediated interactions as a function of $\mu_s$ and $\mu_v$. We normalize $\lambda_{max}$ by its maximum value over the range of $\mu_s$ and $\mu_v$ plotted.}\label{fig:FSlambda}
    \end{subfigure}
    \hfill
    \begin{subfigure}{0.45\textwidth}
        \centering
        \includegraphics[width=\linewidth]{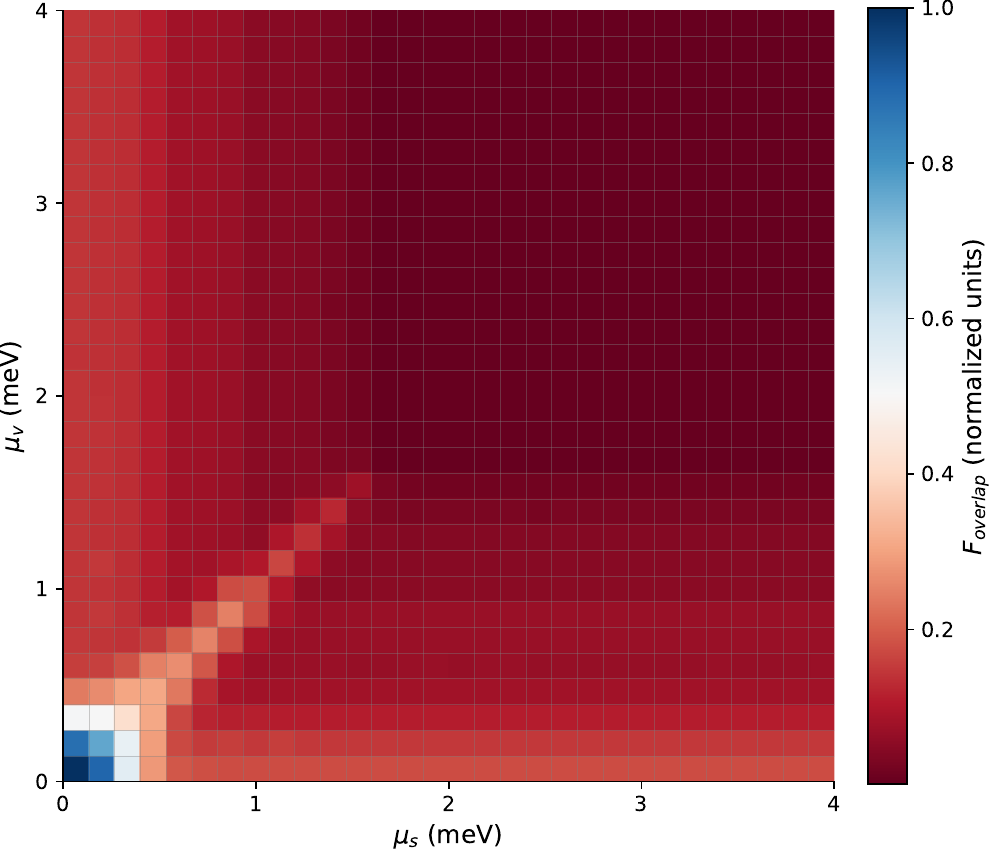}
        \caption{We show the quantity $F_{\text{overlap}}$ as a function of $\mu_s$ and $\mu_v$. We normalize $F_{\text{overlap}}$ by its maximum value over the range of $\mu_s$ and $\mu_v$ plotted.}\label{fig:FSoverlap}
    \end{subfigure}
\end{figure}

When $\mu_v=\mu_s$ in the case of phonons, we observe a re-emergence of an intervalley, spin triplet superconducting order. One can think of this line as originating from the fact that when $\mu_s=\mu_v$ and there is still a partially polarized Fermi surface, a spin-less version of time reversal symmetry is effectively restored, favoring an inter-valley state with no sign change. For sufficiently large $\mu_s$ and $\mu_v$, this line of stability for the intervalley state disappears when the partially polarized Fermi surface becomes partially polarized. As evidenced in the main text, the energetic winner is determined largely by effective overlaps of Fermi surfaces at momenta $\vec{k}$ and $-\vec{k}$ (or $\vec{k}+\vec{q}/2$ and $-\vec{k}+\vec{q}/2$ for the finite $\vec{q}$ case). To demonstrate this, in Figs.~\ref{fig:FSlambda} and \ref{fig:FSoverlap}, we plot the value of the leading eigenvalue for phonon-mediated fluctuations alongside a quantity $F_{\text{overlap}}$ which measure the Fermi surface overlap, which we define as:
\begin{equation}
    F_{\text{overlap}}=\sum_{\alpha\beta}\int d\vec{k} \rho(\epsilon_{\alpha,\vec{k}})\rho(\epsilon_{\beta,-\vec{k}})
\end{equation}
where $\rho_{\alpha,\vec{k}}$ is the normal state density of states at the Fermi level of the band with spin and valley indices labeled by $\alpha$. We observe the general trends in these quantities coincide, supporting the notion that the leading instability is dictated largely by the normal state Fermi-surfaces.

Restricting our solutions to $\vec{q}=0$ may quantitatively shift the phase boundaries we observe. However, we expect the general behavior, namely the behavior of the eigenvalue with respect to the Fermi surface overlap to hold for $q$-shifted Fermi surfaces as well.

\begin{figure}
    \centering
    \includegraphics[width=0.95\textwidth]{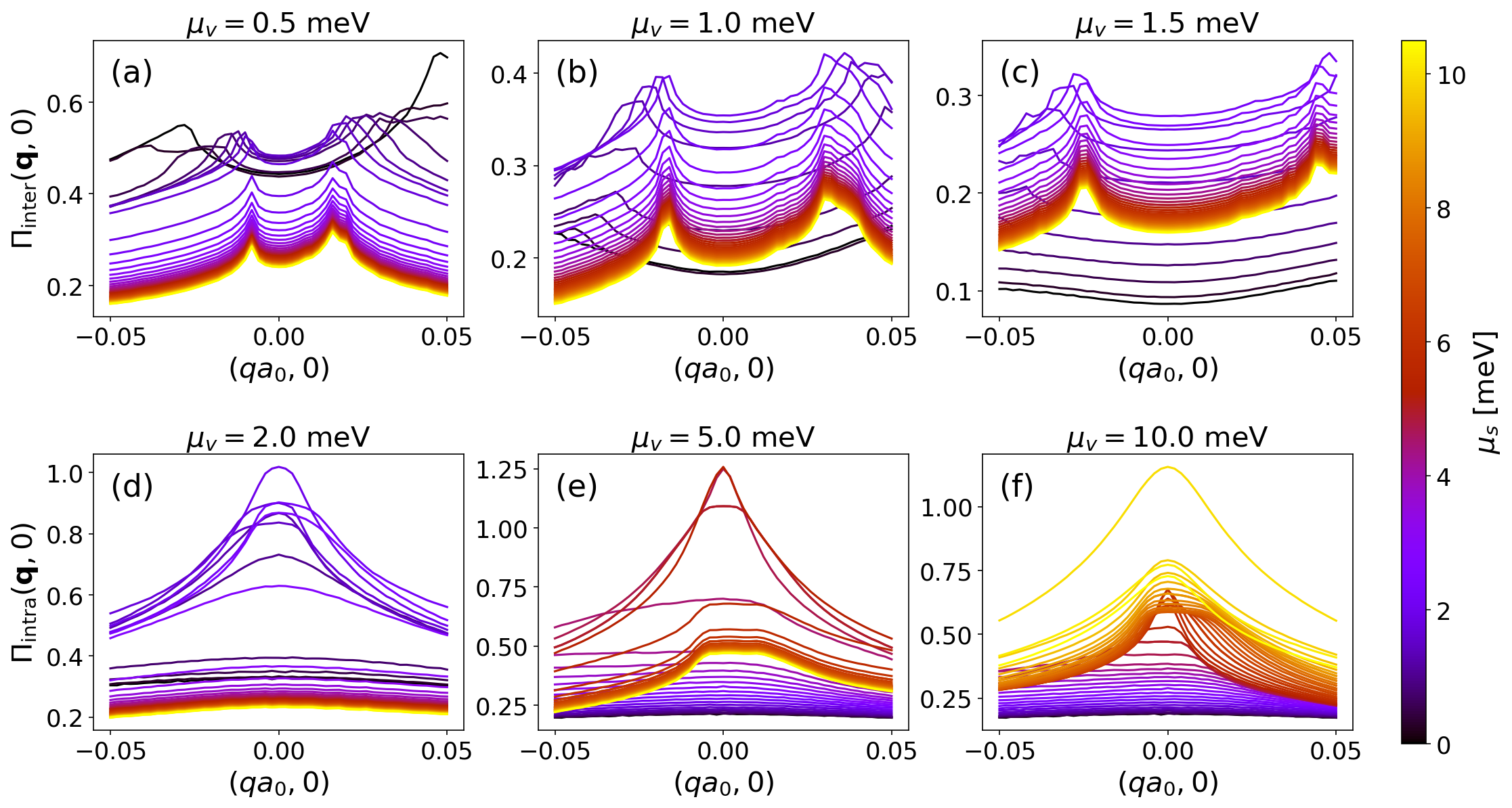}
    \caption{Inter- (panels a-c) and intravalley (panels d-f) pairing bubbles along a momentum cut for different choices of valley ($\mu_v$) and spin ($\mu_s$) polarization.}
    \label{fig: pairing bubbles}
\end{figure}

In Fig.~\ref{fig: pairing bubbles}, we show the behavior of the inter- and intravalley pairing bubbles $\Pi_\mathrm{inter/intra}(\boldsymbol{q},0)$ as functions of the spin polarization $\mu_s$ and along a momentum cut $\boldsymbol{q}=(q,0)$ for several choices of the valley polarization $\mu_v$, and assuming ferromagnetic-fluctuation-mediated pairing.

For smaller valley polarizations (up to $\mu_v\sim 1.8$ meV), the intervalley bubbles are larger the intravalley ones (not shown), independent of the spin polarization, indicating the pairing between different valleys is favored. Conversely, for larger $\mu_v$ intravalley pairing is favored. We note that in the case of intervalley pairing, the bubbles have well-defined peaks at finite momenta that weakly depend on $\mu_s$ and drift away from the $\boldsymbol{q}=(0,0)$ with increasing $\mu_v$.  Differently, the intravalley bubbles exhibit either a single peak at $\boldsymbol{q}=(0,0)$ or a very flat momentum dependence, implying that the corresponding superconducting state will have a small superconducting stiffness, hence strong fluctuations. 

To calculate the bubbles, we have assumed the superconducting order parameter to take the simple form 
\begin{equation}
    \Delta_{\boldsymbol{k},\boldsymbol{q}}=\sum_\sigma\langle c_{\boldsymbol{k}+\frac{\boldsymbol{q}}{2},\sigma}\,c_{-\boldsymbol{k}+\frac{\boldsymbol{q}}{2},\sigma}\rangle\propto f_{\boldsymbol{k}}=\mathrm{exp}\left(-\frac{|\boldsymbol{k}|^2}{k_0^2}\right)\,\frac{k_x+ik_y}{|\boldsymbol{k}|}\,,
\end{equation}
with $k_0=0.1a_0$. The pairing bubbles are then defined as
\begin{equation}
    \begin{split}
        &\Pi_\mathrm{inter}(\boldsymbol{q},0)=\sum_{\sigma,\eta}\int_{\boldsymbol{k}}f_{\boldsymbol{k}}^2\,\frac{f(-\xi_{\boldsymbol{k}+\frac{\boldsymbol{q}}{2},\eta}+\eta\mu_v+\sigma\mu_s)-f(\xi_{-\boldsymbol{k}+\frac{\boldsymbol{q}}{2},-\eta}+\eta\mu_v-\sigma\mu_s)}{\xi_{\boldsymbol{k}+\frac{\boldsymbol{q}}{2},\eta}+\xi_{-\boldsymbol{k}+\frac{\boldsymbol{q}}{2},-\eta}-2\sigma\mu_s}\,,\\
        &\Pi_\mathrm{intra}(\boldsymbol{q},0)=\sum_{\sigma,\eta}\int_{\boldsymbol{k}}f_{\boldsymbol{k}}^2\,\frac{f(-\xi_{\boldsymbol{k}+\frac{\boldsymbol{q}}{2},\eta}+\eta\mu_v+\sigma\mu_s)-f(\xi_{-\boldsymbol{k}+\frac{\boldsymbol{q}}{2},\eta}-\eta\mu_v-\sigma\mu_s)}{\xi_{\boldsymbol{k}+\frac{\boldsymbol{q}}{2},\eta}+\xi_{-\boldsymbol{k}+\frac{\boldsymbol{q}}{2},\eta}-2\eta\mu_v-2\sigma\mu_s}\,,\\
    \end{split}
\end{equation}
where $\xi_{\boldsymbol{k},\eta}=\epsilon_{\boldsymbol{k},\eta}-\mu$, and $f(x)=1/(e^{x/T}+1)$ is the Fermi function. The calculations have been performed at $T=0.01$ meV.

\end{appendix}

\end{document}